\documentclass[11pt]{article}
\usepackage{amssymb}

\usepackage[totalwidth=460truept,totalheight=600truept]{geometry}
\usepackage{epsfig,latexsym}

\def\theequation{\arabic{section}.\arabic{equation}}

\renewcommand{\theequation}{\thesection.\arabic{equation}}
\global\arraycolsep=1truept
\input epsf
\input{tcilatex}
\begin{document}

\hfill \hfill IFUP-TH 2005/07


\vskip 1.4truecm

\begin{center}
{\huge \textbf{Renormalization Of A Class Of}}

\vskip .4truecm

{\huge \textbf{Non-Renormalizable Theories}}

\vskip 1.5truecm

\textsl{Damiano Anselmi}

\textit{Dipartimento di Fisica ``Enrico Fermi'', Universit\`{a} di Pisa, }

\textit{Largo Bruno Pontecorvo 3, I-56127 Pisa, Italy, }

\textit{and INFN, Sezione di Pisa, Italy}

e-mail: anselmi@df.unipi.it
\end{center}

\vskip 2truecm

\begin{center}
\textbf{Abstract}
\end{center}

{\small Certain power-counting non-renormalizable theories, including the
most general self-interacting scalar fields in four and three dimensions and
fermions in two dimensions, have a simplified renormalization structure. For
example, in four-dimensional scalar theories, }$2n${\small \ derivatives of
the fields, }$n>1${\small , do not appear before the }$n${\small th loop. A
new kind of expansion can be defined to treat functions of the fields (but
not of their derivatives) non-perturbatively. I study the conditions under
which these theories can be consistently renormalized with a reduced,
eventually finite, set of independent couplings. I find that in common
models the number of couplings sporadically grows together with the order of
the expansion, but the growth is slow and a reasonably small number of
couplings is sufficient to make predictions up to very high orders. Various
examples are solved explicitly at one and two loops.}

\vskip1truecm

\vfill\eject

\section{Introduction}

\setcounter{equation}{0}

The divergences of power-counting non-renormalizable theories are commonly
subtracted away introducing infinitely many independent couplings in the
theory. In this naive formulation, non-renormalizable theories can be used
only as effective field theories. Although effective field theories are good
for all practical purposes, they do not have much power to suggest new
physics beyond them.

Some features of the Standard Model (such as its large number of independent
parameters) and certain experimental facts (such as the neutrino masses)
point towards a more fundamental theory. Probably, LHC is going to tell us
whether a fundamental Higgs scalar exists or not. If there exists no
right-handed neutrino or no Higgs scalar, it is necessary to consider
theories that include power-counting non-renormalizable interactions and
study their predictive power. The investigation of non-renormalizable
interactions is interesting also for its potential applications to quantum
gravity, although quantum gravity has other problems beyond that of
renormalization. Some physicists consider quantum field theory inadequate or
limited under several respects, and look for approaches beyond quantum field
theory. It is certainly useful to know what can be said on these issues
within quantum field theory, before abandoning it. Probably this
investigation can also suggest the best directions to go beyond quantum
field theory, and clarify if it is really necessary to do so.

The present understanding of non-renormalizable theories is still imperfect.
An improved formulation could be very useful. Some steps in this direction
have been made in ref.s \cite{pap2,pap3}, where finite and quasi-finite
irrelevant deformations of interacting conformal field theories have been
constructed. These non-renormalizable theories contain an infinite number of
lagrangian terms, but only a finite number of independent couplings, and are
renormalized by means of field redefinitions, plus eventually
renormalization constants for the independent couplings. The constructions
are perturbative, which ensures calculability. The purpose of this paper is
to study the properties of another noticeable class of non-renormalizable
theories, to make progress towards the inclusion of more realistic models,
hopefully useful for physics beyond the Standard Model and maybe quantum
gravity.

\bigskip

The first observation of this paper is that certain theories have a
simplified renormalization structure, even if they contain power-counting
non-renormalizable interactions. Using this property it is possible to treat
functions of the fields (but not of their derivatives) non-perturbatively.
Non-analytic dependencies are allowed also.

More precisely, there is a relation between the loop expansion and the
expansion in the derivatives of the fields. For example, in four-dimensional
scalar theories, terms with $2n$ derivatives of the fields, $n>1$, do not
appear before the $n^{\text{th}}$ loop. Schematically, the renormalizable
lagrangian has the form 
\begin{equation}
\mathcal{L}=\frac{1}{2}(\partial \varphi )^{2}+\sum_{n=0}^{\infty
}\!\!\!\!\!\!\left. \phantom{{a\over a}}\right. ^{\prime }\hbar
^{n}[\partial ^{2n}]V_{n}(\varphi ,\hbar ),  \label{master}
\end{equation}
where $[\partial ^{2n}]$ is a symbolic notation to denote $2n$ variously
distributed derivatives of the fields $\varphi $, contracted in all possible
ways. The prime over the sum means that only inequivalent contractions are
considered. Equivalent contractions are those that differ by total
derivatives and field redefinitions.

The functions $V_{n}(\varphi ,\hbar )$ need not be polynomial, nor analytic,
and can be treated non-per\-tur\-ba\-ti\-ve\-ly, working directly on the
quantum action $\Gamma [\varphi ]$, instead of working on correlation
functions. The background field method \cite{backfmeth} is pratically
convenient for this purpose.

The quantum action $\Gamma $ can be expanded in powers 
\begin{equation}
\left( \frac{E}{M_{P\text{eff}}^{{}}}\right) ^{m}\left( \frac{\alpha M_{P%
\text{eff}}^{2}}{E^{2}}\right) ^{n},  \label{yt}
\end{equation}
where $M_{P\text{eff}}$ is an ``effective Planck mass'', $E$ is the energy
and $\alpha $ collectively denotes the marginal couplings of the theory.
Every quantity is calculable with a finite number of steps at each order of
this expansion, which means for energies $E$ in the range $\alpha M_{P\text{%
eff}}^{2}\ll E^{2}\ll M_{P\text{eff}}^{2}$. It is not necessary to assume
that the field $\varphi $ is small with respect to $M_{P\text{eff}}$, but
just that its momentum $p\sim E$ is. More common expansions are meaningful
in the presence of a mass gap, but the expansion (\ref{yt}) is meaningful
also in the absence of a mass gap.

These tools can be used to study the predictive power of the theories (\ref
{master}), in particular the conditions under which the divergences can be
renormalized with a reduced, eventually finite, set of independent
couplings. Start from a classical potential, e.g. $V_{\text{cl}}(\varphi
,0)=\lambda _{0}\varphi ^{4}/4!+\lambda _{2}\varphi ^{6}/6!$ and look for a
completion (\ref{expa}) that is stable under renormalization and possibly
depends only on the couplings contained in $V_{\text{cl}}$. Imposing
consistency with renormalization and other simple requirements differential
or integro-differential equations for the functions $V_{n}$ are obtained.
The $V_{n}$-equations determine a ``corrected classical potential'' $%
V_{0}(\varphi ,0)$ from the ``initial conditions'' $V_{\text{cl}}(\varphi
,0) $, and the functions $V_{n}(\varphi ,\hbar )$s, $n\geq 0$, from $%
V_{0}(\varphi ,0)$. The non-perturbative solutions are regular even where
the expansions in powers of the fields do not converge.

I find that, in general, the initial conditions $V_{\text{cl}}$ are not
sufficient to solve the $V_{n}$-equations unambiguously and new independent
couplings need to be introduced along the way. When $V_{\text{cl}}$ is
analytic, the number of new couplings grows sporadically with the order of
the perturbative expansion, but the growth is so slow that a reasonably
small number of couplings is sufficient to make predictions up to very high
orders. In the end the complete lagrangian (\ref{master}) contains
infinitely many couplings, although each $V_{n}(\varphi ,\hbar )$ depends on
finitely many. Instead, a class of non-analytic theories, e.g. those defined
by $V_{\text{cl}}(\varphi ,0)=\lambda _{0}\varphi ^{4}/4!+\lambda
_{r}\varphi ^{r+4}/(r+4)!$ with $r^{2}$ irrational, are renormalized with a
finite number of independent couplings, typically just those contained in $%
V_{\text{cl}}$.

\bigskip

The study of quantum field theory beyond power counting has attracted
interest for decades, motivated by low-energy QCD, quantum gravity and the
search for new physics beyond the Standard Model. Some non-renormalizable
theories can be given sense thanks to \textit{ad hoc} procedures, such as
the large N expansion used to construct three-dimensional four-fermion
theories \cite{parisi} and sigma models \cite{arefeva2}. A more general
proposal for non-renormalizable theories is Weinberg's asymptotic safety 
\cite{wein}, where the theory is assumed to have an interacting fixed point
in the ultraviolet, with a finite-dimensional critical surface. The RG flow
tends to the fixed point only if the irrelevant couplings are appropriately
fine-tuned. This fine-tuning leaves a finite number of arbitrary parameters.
It is difficult to prove the existence of an interacting UV fixed point with
the usual perturbative techniques, so alternative methods have been used,
such as the lattice and the exact renormalization-group (ERG) approach.
Asymptotic safety in the ERG framework has been recently studied for gravity 
\cite{reuter} and the Higgs sector of the Standard Model \cite{wette}.

The main difference between $a$) the ERG asymptotic-safety approach and $b$)
the approach of this paper and ref.s \cite{pap2,pap3} is that $a$) is
essentially non-perturbative and relies on the existence of a UV fixed
point, while $b$) is perturbative or partially non-perturbative and does not
assume the existence of a UV fixed point. In approach $a$) the attention is
focused on the overall scale dependence, so the RG\ flow is defined varying
the overall energy with respect to the dimensionful couplings and the
dynamical scale $\mu $. Instead, in approach $b$) the attention is focused
on the role of divergences, so the RG flow is defined varying the
renormalization point $\mu $ with respect to the the overall energy and the
dimensionful couplings. In particular, the finite non-renormalizable
theories of \cite{pap2,pap3} are fixed points of the RG flow in the $b$%
)-sense, but not fixed points of the RG\ flow in the $a$)-sense, since they
contain an explicit scale, the Planck mass. The two approaches $a$) and $b$)
pursue similar objectives starting from different assumptions and often
using rather different techniques, so it would be extremely interesting to
understand how they can be merged, to gain the maximum of information from
both. For example, the partially non-perturbative techniques developed here
might be useful to search for a UV fixed point in the $b$)-sense in some
cases.

Techniques to reduce the number of independent couplings consistently with
renormalization have been first studied by Zimmermann \cite{zimme,oheme} in
the realm of power-counting renormalizable theories. The reduction of
couplings in non-renormalizable theories has rather different properties,
that deserve to be studied apart. Section 4 contains the knowledge that is
useful for the purposes of this paper, while a systematic treatment of this
issue is left to separate publications.

An analysis of non-renormalizable theories beyond power counting was
performed by Atance and Cortes in ref.s \cite{cortes1,cortes2} within the
usual expansion in powers of the fields and their derivatives, by Kubo and
Nunami \cite{giap} and Halpern and Huang \cite{halpern} using the Wilsonian
approach. Here, besides the progress made in the directions outlined above,
some claims of ref. \cite{cortes1} are corrected, with particular attention
to the number of independent couplings, that in general grows sporadically
together with the order of the perturbative expansion.

\medskip

The issues treated here are technically involved. To rationalize the
presentation I first derive the general properties and later give results of
explicit computations in various models, at one and two loops, using the
dimensional-regularization technique in the Euclidean framework. Such models
do not have immediate realistic applications, but the investigation of their
renormalization properties is instructive, because they provide a laboratory
and a fruitful arena to test ideas that might inspire the research about
physics beyond the Standard Model and the renormalization of quantum gravity.

The paper is organized as follows. In section 2 I prove the
renormalizability of lagrangian structures such as (\ref{master}) and study
various properties of those, in particular their simplified dependence on
the marginal and irrelevant couplings. In section 3 I derive the
perturbative expansion that is compatible with (\ref{master}). Section 4
contains the main properties of the reduction of couplings in
non-renormalizable theories. In section 5 I solve the $\varphi ^{4}+\varphi
^{6}$ theory in four dimensions, at one and two loops. In section 6 I
consider the $\varphi ^{4}+\varphi ^{m+4}$ theory in four dimensions and
study under which conditions the divergences can be renormalized with a
reduced, eventually finite, number of independent couplings to all orders in
the perturbative expansion. In section 7 I study a case where a new coupling
appears already at the tree level: the $\varphi ^{4}+\varphi ^{5}$ theory in
four dimensions. Section 8 contains the conclusions.

\section{Structure of the renormalized lagrangian}

\setcounter{equation}{0}

In this section I prove that certain theories have simplified
renormalization structures and study how they depend on the marginal and
irrelevant couplings. This class of theories includes scalar fields in four
and three spacetime dimensions, fermions in two dimensions and
scalar-fermion theories coupled in a peculiar way. The rationalized
lagrangian of scalar fields in four dimensions is written schematically in
equation (\ref{master}). Here $2n$ derivatives of the fields, $n>1$, do not
appear before the $n^{\text{th}}$ loop. The functions $V_{n}(\varphi ,\hbar
) $s are for the moment generic functions, that renormalize with the rules
worked out below. In three-dimensional scalar theories and two-dimensional
fermion theories $n$ derivatives of the fields do not appear before the $n^{%
\text{th}}$ loop, for $n>2$ and $n>1$, respectively. In coupled
scalar-fermion theories some non-derivative vertices, such as the Yukawa
coupling, are multiplied by powers of $\hbar $.

The results of this section do not ensure that the divergences are removed
with a finite number of independent couplings. Nevertheless, they classify
the divergences in a convenient way, that is useful both for applications to
effective field theory and to study the conditions under which a reduced,
eventually finite, set of independent couplings are sufficient to
renormalize the divergences.

\bigskip

\textbf{Renormalization structure}. Consider a generic non-renormalizable
theory, described by a lagrangian $\mathcal{L}$. Let $\mathcal{V}$ denote
the set of its vertices. Typically, these are infinitely many. Let $%
\widetilde{\mathcal{L}}$ denote a ``reduction'' of the theory, namely a
lagrangian that contains only some finite subset $\widetilde{\mathcal{V}}$
of vertices. In general, the reduced theory $\widetilde{\mathcal{L}}$ is not
stable under renormalization, which means that not all counterterms
generated by the Feynman diagrams constructed with the vertices of the
subset $\widetilde{\mathcal{V}}$ are contained in $\widetilde{\mathcal{V}}$.
Then, new vertices have to be added to $\widetilde{\mathcal{L}}$, multiplied
by independent couplings. Call the extended lagrangian $\widetilde{\mathcal{L%
}}_{\text{ext}}$ and the extended set of vertices $\widetilde{\mathcal{V}}_{%
\text{ext}}$. The renormalization constants of the couplings contained in $%
\widetilde{\mathcal{L}}_{\text{ext}}-\widetilde{\mathcal{L}}$ cancel the new
divergences, those of type $\widetilde{\mathcal{V}}_{\text{ext}}\backslash 
\widetilde{\mathcal{V}}$. Typically $\widetilde{\mathcal{L}}_{\text{ext}}$
is itself unstable, and the extension has to be iterated, possibly
infinitely many times. This iteration defines a theory $\mathcal{L}_{s}$
that is certainly stable, but in general contains all possible vertices (so
it coincides with $\mathcal{L}$).

In this iterative procedure, the new vertices are normally introduced at the
tree level. Then the stable theory $\mathcal{L}_{s}$ contains all possible
vertices already at the tree level. However, in some models it is possible
to rationalize the renormalization structure, in the following sense. Assume
that the Feynman diagrams constructed with the vertices of the reduced
lagrangian $\widetilde{\mathcal{L}}$ generate a divergence proportional to a
certain vertex $v$ only at order $\hbar ^{n}$. Then this divergence is local
and a simple pole \cite{collins}, so it has the form 
\begin{equation}
\frac{1}{\varepsilon }\hbar ^{n}v  \label{div}
\end{equation}
where $\varepsilon =D-d$, $D$ the physical spacetime dimension and $d$ is
its continuation in dimensional regularization. Include a vertex $\hbar
^{n-1}v$ in the extended theory $\widetilde{\mathcal{L}}_{\text{ext}}%
\mathcal{\ }$and cancel the divergence (\ref{div}) with the $\mathcal{O}%
(\hbar )$-contribution to the renormalization constant $Z_{v}$ of $v$. If
also the extended theory does not generate (\ref{div}) before the order $%
\hbar ^{n}$, then the extension is stable. In this case, it is unnecessary
to introduce the vertex $v$ at orders lower than $\hbar ^{n-1}$. It is said
that the coupling $v$ does not appear before the order $\hbar ^{n-1}$.

In this section I study the models that admit this kind of stable
renormalization structure. In the rest of the paper I use this reduction as
a tool to study a further, more important, reduction, namely the reduction
of couplings.

\bigskip

More precisely, suppose that the theory contains vertices that do not
contribute at the tree level, because they are multiplied by powers of $%
\hbar $. Call $\mathcal{L}_{n}$ the lagrangian truncated up to the order $%
\hbar ^{n}$ included and let $\mathcal{V}_{n}$ denote the set of vertices of 
$\mathcal{L}_{n}$. The set $\{\mathcal{V}_{n}\}$ is a \textit{%
renormalization structure}. Obviously, $\mathcal{V}_{n}\subseteq \mathcal{V}%
_{n+1}$. Let $\mathcal{C}_{n+1}$ denote the counterterms of order up to $%
\hbar ^{n+1}$. The counterterms $\mathcal{C}_{n+1}$ are originated by
Feynman diagrams constructed with the vertices $\mathcal{V}_{n}$. Observe
that the $\hbar $-powers that multiply a diagram are partially due to the
number of loops and partially due to the $\hbar $-powers that multiply the
vertices. The structure $\{\mathcal{V}_{n}\}$ is stable under
renormalization if $\mathcal{C}_{n+1}\subseteq \mathcal{V}_{n}$ for every $n$%
.

In the example (\ref{master}) $\mathcal{V}_{n}$ contains vertices with at
most $2n$ derivatives and an arbitrary number of legs. Below I prove that
the structure (\ref{master}) is stable under renormalization.

\bigskip

Consider a classical theory of scalar fields interacting by means of the
lagrangian 
\begin{equation}
\mathcal{L}=\frac{1}{2}(\partial _{\mu }\varphi )^{2}+V(\varphi )  \label{th}
\end{equation}
in four spacetime dimensions. For the moment I am interested only in the UV
divergences of the quantum theory, and their renormalization. Then it is
consistent to treat the mass term, if present, as a (two-leg) vertex of $V$,
the propagator being just $1/k^{2}$. To avoid IR\ problems in the
intermediate steps, it is convenient to calculate the UV divergences of
Feynman diagrams with a deformed propagator $1/(k^{2}+\delta ^{2})$ and let $%
\delta $ tend to zero at the end. The limit is smooth, since the divergent
parts of Feynman diagrams depend polynomially on $\delta $. The tadpoles are
loops with a single vertex and vanish identically. However, note that loops
with at least two vertices are not tadpoles (even if one of the vertices is
a two-leg ``mass'' vertex) and do give a non-trivial divergent contribution,
which can be calculated at $\delta \neq 0$.

Now I prove that, taking the quantum corrections into account, the complete
renormalizable lagrangian has the form 
\begin{equation}
\mathcal{L}=\frac{1}{2}(\partial _{\mu }\varphi )^{2}+\sum_{n=0}^{\infty
}\!\!\!\!\!\!\left. \phantom{{a\over a}}\right. ^{\prime }\hbar
^{n}[\partial ^{2n}]V_{n}(\varphi ,\hbar ).~  \label{expa}
\end{equation}
Obviously, there is a finite number of independent ways to distribute and
contract the partial derivatives for every $n$ and the symbolic notation
understands also the sum over them. The prime over the sum means that terms
differing by total derivatives are equivalent and that contractions
factorizing a $\Box \varphi $ are ignored. Indeed, the terms proportional to 
$\Box \varphi $ can be converted into terms of other types by means of a
field redefinition. In particular, note that all terms with $n=1$ in (\ref
{expa}) belong to such class, so they do not contribute to the sum.

The tree lagrangian coincides with (\ref{th}), where $V(\varphi
)=V_{0}(\varphi ,0)$. The functions $V_{n}(\varphi ,\hbar )$ are power
series in $\hbar $ and can be understood as field-dependent ``coupling
constants'' of the theory. In particular, the renormalization is achieved by
means of a field redefinition 
\begin{equation}
\varphi \rightarrow \phi (\varphi )=\varphi +\hbar \sum_{n=0}^{\infty }\hbar
^{n}[\partial ^{2n}]F_{n}(\varphi ,\hbar ,\varepsilon )\text{,}\qquad
F_{n}(\varphi ,\hbar ,\varepsilon )=\sum_{m=1}^{\infty }\frac{\hbar ^{m}}{%
\varepsilon ^{P^{\prime }(m)}}F_{n,m}(\varphi ,\varepsilon ),  \label{fren}
\end{equation}
plus redefinitions 
\begin{equation}
V_{n}(\varphi ,\hbar )\rightarrow V_{\mathrm{R}n}(\phi ,\hbar ,\varepsilon
)=V_{n}(\phi ,\hbar )+\sum_{m=1}^{\infty }\frac{\hbar ^{m}}{\varepsilon
^{P(m)}}V_{n,m}(\phi ,\varepsilon )\text{,}  \label{vren}
\end{equation}
where $P(m)\leq m$, $P^{\prime }(m)\leq m$ and $V_{n,m}(\phi ,\varepsilon )$%
, $F_{n,m}(\varphi ,\varepsilon )$ are analytic functions of $\varepsilon $.
The renormalized lagrangian reads 
\begin{equation}
\mathcal{L}_{\mathrm{R}}(\varphi )=\frac{1}{2}(\partial _{\mu }\phi
)^{2}+\sum_{n=0}^{\infty }\!\!\!\!\!\!\left. \phantom{{a\over a}}\right.
^{\prime }\hbar ^{n}[\partial ^{2n}]V_{\mathrm{R}n}(\phi ,\hbar ,\varepsilon
),  \label{renexpa}
\end{equation}
or, equivalently, 
\begin{equation}
\mathcal{L}_{\mathrm{R}}(\varphi )=\frac{1}{2}(\partial _{\mu }\varphi
)^{2}+\sum_{n=0}^{\infty }\hbar ^{n}[\partial ^{2n}]\widetilde{V}_{\mathrm{R}%
n}(\varphi ,\hbar ,\varepsilon ),  \label{renexpaext}
\end{equation}
where now the sum includes contractions of indices factorizing $\Box \varphi 
$, up to total derivatives, and the functions $\widetilde{V}_{\mathrm{R}n}$
have the same structure as the functions $V_{\mathrm{R}n}$ of (\ref{vren}).
All functions of $\hbar $ written in this paper are meant to be power series
in $\hbar $.

With the exception of the free kinetic term, each finite term of (\ref{expa}%
), (\ref{renexpa}) and (\ref{renexpaext}) satisfies the key inequality 
\begin{equation}
\omega _{\partial }\leq 2\omega _{\hbar },  \label{key}
\end{equation}
where $\omega _{\partial }$ is the number of derivatives and $\omega _{\hbar
}$ is the power of $\hbar $. All counterterms satisfy 
\begin{equation}
\omega _{\partial }\leq 2(\omega _{\hbar }-1),  \label{key2}
\end{equation}
while the counterterms contained in the field redefinition (\ref{fren})
satisfy 
\begin{equation}
\omega _{\partial }\leq 2(\omega _{\hbar }-2).  \label{key3}
\end{equation}
Observe that a field redefinition that fulfills (\ref{key3})\ preserves (\ref
{key}) and (\ref{key2}).

\bigskip

\textbf{Proof of renormalizability.} Now I prove that the lagrangian (\ref
{expa}) is renormalizable in the form (\ref{renexpa}).

Let $G$ denote a Feynman diagram with $L$ loops, $V$ vertices, $I$ internal
legs and $v_{n}$ vertices with $2n$ derivatives. The superficial degree of
divergence is 
\begin{equation}
\omega (G)=-2I+4L+2\sum_{n}nv_{n}.  \label{a1}
\end{equation}
Using the formula $I=L+V-1,$ that relates the number of internal legs to the
number of loops and the number of vertices, (\ref{a1}) becomes 
\begin{equation}
\omega (G)=2\left( L+\sum_{n}nv_{n}-V+1\right) .  \label{obaoba}
\end{equation}
Formula (\ref{expa}) shows that every derivative vertex carries a power of $%
\hbar $. The total $\hbar $-power of the diagram $G$ is denoted with $L_{e}$
and called ``effective number of loops'', to distinguish it from the true
number of loops $L$. Clearly, $L_{e}$ is equal to $L$ plus the powers of $%
\hbar $ attached to the vertices. Each vertex with $2n$ derivatives carries
at least $n$ powers of $\hbar $. Write 
\begin{equation}
L_{e}=L+\sum_{n}nv_{n}+\Delta ,\qquad \Delta \geq 0.  \label{inequa}
\end{equation}
Then 
\begin{equation}
\omega (G)=2(L_{e}+1-\Delta -V)\leq 2(L_{e}-V+1).  \label{degree}
\end{equation}
The last inequality shows that the degree of divergence decreases when the
number of vertices increases, therefore the number of divergent diagrams of
order $\hbar ^{L_{e}}$ is finite. Since the mass term is treated as a
vertex, the tadpoles vanish, so the number $V$ of vertices can be restricted
to $V\geq 2$, which implies 
\begin{equation}
\omega (G)\leq 2(L_{e}-1).  \label{iu}
\end{equation}
In particular, the counterterms satisfy (\ref{key2}).

Now, proceed inductively in $L_{e}$. Assume that the theory is renormalized
by (\ref{renexpa}) and (\ref{fren}) up to the order $\hbar ^{L_{e}-1}$
included. Then the theorem of locality of counterterms \cite{collins}
ensures that the $\mathcal{O}(\hbar ^{L_{e}})$-counterterms are local. Thus
the counterterms due to the diagram $G$ have the form 
\[
\frac{1}{\varepsilon ^{P(G)}}\hbar ^{L_{e}}[\partial ^{\omega
(G)}]H_{G}(\varphi ,\varepsilon )=\hbar ^{\omega (G)/2}[\partial ^{\omega
(G)}]\frac{\hbar ^{L_{e}-\omega (G)/2}}{\varepsilon ^{P(G)}}H_{G}(\varphi
,\varepsilon ), 
\]
where $P(G)$ is the maximal pole of the diagram and $H_{G}(\varphi
,\varepsilon )$ is a certain function of $\varphi $, analytic in $%
\varepsilon $. The theorem proved in the appendix ensures that $P(G)\leq V-1$%
. Then, associating a factor $\hbar ^{\omega (G)/2}$ with the derivatives as
in (\ref{expa}), the remaining power of $\hbar $ satisfies 
\[
L_{e}-\frac{1}{2}\omega (G)=V-1+\Delta \geq V-1\geq P(G)\text{,} 
\]
namely the counterterms have the form specified by (\ref{fren}), (\ref{vren}%
) and (\ref{renexpa}).

In total, using (\ref{iu}), the $\mathcal{O}(\hbar ^{L_{e}})$-counterterms
read 
\begin{equation}
\Delta \mathcal{L}_{L_{e}}=\hbar ^{L_{e}}\sum_{k=0}^{L_{e}-1}\frac{1}{%
\varepsilon ^{p(k)}}[\partial ^{2k}]H_{L_{e},k}(\varphi ,\varepsilon ),
\label{count}
\end{equation}
where $p(k)\leq $ $L_{e}-k$ and $H_{L_{e},k}(\varphi ,\varepsilon )$ are
certain analytic functions of $\varepsilon $.

Consider the counterterms that factorize a $\Box \varphi $, which have the
form 
\[
\Delta _{1}\mathcal{L}_{L_{e}}=\hbar ^{L_{e}}\Box \varphi
\sum_{k=1}^{L_{e}-1}\frac{1}{\varepsilon ^{p(k)}}[\partial ^{2(k-1)}%
]K_{L_{e},k}(\varphi ,\varepsilon ). 
\]
Such counterterms can be reabsorbed into a field redefinition 
\begin{equation}
\delta \varphi =-\hbar ^{L_{e}}\sum_{k=1}^{L_{e}-1}\frac{1}{\varepsilon
^{p(k)}}[\partial ^{2(k-1)}]K_{L_{e},k}(\varphi ,\varepsilon )  \label{fref}
\end{equation}
of type (\ref{fren}), since it satisfies (\ref{key3}) and $p(k)\leq $ $%
L_{e}-k$. The divergent terms generated by this field redefinition contain
all orders in $\hbar $, starting from $\hbar ^{L_{e}}$. They respect the
inequality (\ref{key2}) and the structure (\ref{renexpaext}): 
\begin{equation}
\mathcal{L}_{\mathrm{R}}[\phi (\varphi )+\delta \varphi ]-\mathcal{L}_{%
\mathrm{R}}[\phi (\varphi )]=-\delta \varphi \Box \phi -\frac{1}{2}\delta
\varphi \Box \delta \varphi +\sum_{j=1}^{\infty }\frac{1}{j!}(\delta \varphi
)^{j}\frac{\partial ^{j}}{\partial \phi ^{j}}\sum_{n=0}^{\infty
}\!\!\!\!\!\!\left. \phantom{{a\over a}}\right. ^{\prime }\hbar
^{n}[\partial ^{2n}]V_{\mathrm{R}n}(\phi ,\hbar ,\varepsilon ).  \label{bodo}
\end{equation}
The orders higher than $\hbar ^{L_{e}}$ can be ignored at this step of the
inductive procedure. In the subsequent steps they will be subtracted away
just as they come (i.e. as ``diagrams'' with no true loop, $L=0$), because
they are local, divergent and satisfy (\ref{key2}),(\ref{iu}), as the
counterterms with $L>0$. The counterterms of order $\hbar ^{L_{e}}$
contained on the left-hand side of (\ref{bodo}) are 
\begin{equation}
\delta \mathcal{L}\equiv \Delta _{1}\mathcal{L}_{L_{e}}+\frac{\partial V_{%
\mathrm{R}0}}{\partial \varphi }\delta \varphi .  \label{osto}
\end{equation}
It is immediate to prove that the difference $\Delta \mathcal{L}%
_{L_{e}}-\delta \mathcal{L}$ has the same form as (\ref{count}), 
\begin{equation}
\hbar ^{L_{e}}\sum_{k=0}^{L_{e}-1}\frac{1}{\varepsilon ^{p(k)}}[\partial
^{2k}]H_{L_{e},k}^{\prime }(\varphi ,\varepsilon ),  \label{count2}
\end{equation}
for certain modified functions $H^{\prime }$, but carries fewer factors $%
\Box \varphi $ than (\ref{count}). In general (\ref{count2}) is not in the
final irreducible form, yet. Other terms factorising $\Box \varphi $ can be
brought by the last term of (\ref{osto}), due to factors $\Box \varphi $
inside $\delta \varphi $ (\ref{fref}), or, equivalently, factors $(\Box
\varphi )^{n}$, $n>1$, in (\ref{count}). Obviously, factors $(\Box \varphi
)^{n}$ in (\ref{count}) generate factors $(\Box \varphi )^{n-1}$ in (\ref
{fren}) and therefore (\ref{count2}). Thus, repeating the field-redefinition
procedure (\ref{count})$\rightarrow $(\ref{count2}) a finite number of
times, the surviving $\mathcal{O}(\hbar ^{L_{e}})$-counterterms acquire the
irreducible form 
\begin{equation}
\hbar ^{L_{e}}\sum_{k=0}^{L_{e}-1}\!\!\!\!\!\!\left. \phantom{{a\over a}}%
\right. ^{\prime }\frac{1}{\varepsilon ^{p(k)}}[\partial ^{2k}]\widetilde{H}%
_{L_{e},k}(\varphi ,\varepsilon ).  \label{sta}
\end{equation}
The irreducible counterterms (\ref{sta}) are reabsorbed into redefinitions 
\begin{equation}
V_{\mathrm{R}n}(\phi ,\hbar ,\varepsilon )\rightarrow V_{\mathrm{R}n}(\phi
,\hbar ,\varepsilon )+\frac{\hbar ^{L_{e}-n}}{\varepsilon ^{p(n)}}\widetilde{%
H}_{L_{e},n}(\phi ,\varepsilon ),\qquad \text{for }n=0\text{ and }1<n<L_{e},
\label{autoreno}
\end{equation}
which have the form (\ref{vren}). This concludes the proof that the
structure (\ref{renexpa}) is preserved by renormalization.

\bigskip

\textbf{Dependence on the marginal and irrelevant couplings.} Now, suppose
that the theory (\ref{th}) has no relevant coupling, a marginal coupling $%
\alpha $ and irrelevant couplings $\lambda $. Let $M_{P}$ denote a reference
scale for the irrelevant couplings, which I call the ``Planck mass''. Now I
prove that the renormalization structure (\ref{expa}) is compatible with the
following $\alpha $-$M_{P}$-$\hbar $-dependence: 
\begin{equation}
V_{n}(\alpha ,\varphi ,M_{P},\hbar )\equiv \alpha M_{P}^{4-2n}W_{n}(\hbar
\alpha ,\chi ),\qquad W_{n}(\hbar \alpha ,\chi )=\sum_{k=0}^{\infty }(\hbar
\alpha )^{k}W_{n,k}(\chi ),  \label{vnform0}
\end{equation}
where $\chi =\varphi /M_{P}$. Then the renormalization structure reads 
\begin{equation}
\mathcal{L}=\frac{M_{P}^{2}}{2}(\partial _{\mu }\chi
)^{2}+\sum_{n=0}^{\infty }\!\!\!\!\!\!\left. \phantom{{a\over a}}\right.
^{\prime }\hbar ^{n}\alpha M_{P}^{4-2n}[\partial ^{2n}]W_{n}(\hbar \alpha
,\chi ),  \label{imprexpa}
\end{equation}
whose terms satisfy 
\begin{equation}
\omega _{\partial }+2(\omega _{\alpha }-\omega _{\hbar })=2,  \label{k}
\end{equation}
and, with the exception of the free kinetic term, $\omega _{\alpha }\geq 1$, 
$\omega _{\alpha }$ denoting the power of $\alpha $. The renormalized
lagrangian is obtained from (\ref{imprexpa}) with the replacements 
\begin{equation}
W_{n}(\hbar \alpha ,\chi )\rightarrow W_{\mathrm{R}n}(\hbar \alpha ,\chi
,\varepsilon )=W_{n}(\hbar \alpha ,\chi )+\sum_{m=1}^{\infty }\frac{(\hbar
\alpha )^{m}}{\varepsilon ^{P(m)}}W_{n,m}(\chi ,\varepsilon ),  \label{conta}
\end{equation}
and 
\begin{equation}
\chi \rightarrow \chi +\hbar \alpha \sum_{n=0}^{\infty }\hbar
^{n}M_{P}^{-2n}[\partial ^{2n}]F_{n}(\hbar \alpha ,\chi ,\varepsilon
),\qquad F_{n}(\hbar \alpha ,\chi ,\varepsilon )=\sum_{m=1}^{\infty }\frac{%
(\hbar \alpha )^{m}}{\varepsilon ^{P^{\prime }(m)}}F_{n,m}(\chi ,\varepsilon
),  \label{frefa}
\end{equation}
with $P(m)\leq m$, $P^{\prime }(m)\leq m$ and $W_{n,m}$, $F_{n,m}$ analytic
functions of $\varepsilon $. The divergent contributions of (\ref{conta})
satisfy (\ref{k}) with $\omega _{\alpha }\geq 2$, while those of (\ref{frefa}%
) satisfy $\omega _{\partial }+2(\omega _{\alpha }-\omega _{\hbar })=0$ with 
$\omega _{\alpha }\geq 2$. Obviously, a lagrangian that satisfies (\ref{k})
is preserved by a field redefinition (\ref{frefa}).

Consider a diagram $G$ of order $\hbar ^{L_{e}}$ with $L$ loops, $I$
internal legs and $w_{n,k}$ vertices of type $W_{n,k}$. Let $%
v_{n}=\sum_{k}w_{n,k}$, $V=\sum_{n}v_{n}$. The total power of $\hbar $
attached to the vertices is 
\begin{equation}
L_{e}-L=\sum_{n,k}(k+n)w_{n,k}  \label{usti}
\end{equation}
and the total power of $\alpha $ multiplying the diagram is, using the
definition (\ref{inequa}), 
\begin{equation}
\sum_{n,k}(k+1)w_{n,k}=\sum_{n,k}(k+n-n)w_{n,k}+V=L_{e}-L-%
\sum_{n}nv_{n}+V=V+\Delta \text{.}  \label{obolo}
\end{equation}
The structure of the counterterms is therefore 
\[
\Delta \mathcal{L}_{L_{e}}=\hbar ^{L_{e}}\sum_{G}\frac{1}{\varepsilon ^{P(G)}%
}\alpha ^{V+\Delta }M_{P}^{4V-2\sum_{n}nv_{n}-2I}[\partial ^{\omega
(G)}]H_{G}(\chi ,\varepsilon ), 
\]
with $P(G)\leq V-1$. The power $M_{P}^{-2I}$ is due to the propagators $%
\hbar /(p^{2}M_{P}^{2})$ of the rescaled fields $\chi $. Using (\ref{obaoba}%
) and (\ref{obolo}) we have 
\begin{equation}
\Delta \mathcal{L}_{L_{e}}=\sum_{G}\hbar ^{N(G)}\alpha M_{P}^{4-2N(G)}\frac{%
(\hbar \alpha )^{L_{e}-N(G)}}{\varepsilon ^{P(G)}}[\partial ^{2N(G)}%
]H_{G}(\chi ,\varepsilon ),  \label{tem}
\end{equation}
where $N(G)=L+\sum_{n}nv_{n}-V+1=\omega (G)/2$. Observe that, from (\ref{iu}%
), $L_{e}-N(G)\geq 1$, so (\ref{tem}) satisfies (\ref{k}) with $\omega
_{\alpha }\geq 2$. Moreover, $P(G)\leq L_{e}-N(G)$, in agreement with (\ref
{conta}). The terms proportional to the field equations are reabsorbed with
field redefinitions of the form (\ref{frefa}). After the field redefinitions
the renormalization of $W_{N(G)}$ reads 
\[
W_{N(G)}(\hbar \alpha ,\chi )\rightarrow W_{N(G)}(\hbar \alpha ,\chi )+\frac{%
(\hbar \alpha )^{L_{e}-N(G)}}{\varepsilon ^{P(G)}}\widetilde{H}_{G}(\chi
,\varepsilon ). 
\]
Thus the structure (\ref{vnform0})-(\ref{imprexpa}) is preserved by
renormalization.

\bigskip

\textbf{Extensions to other theories}. The renormalization structure (\ref
{expa})\ is peculiar of scalar field theories in four dimensions. No similar
renormalization structure works in higher dimensions, but the results proved
so far do admit generalizations to theories in lower dimensions and to other
four-dimensional theories.

For example, it is immediate to prove that scalar field theories in three
spacetime dimensions admit the renormalization structure 
\begin{equation}
\mathcal{L}_{D=3}=\frac{1}{2}(\partial _{\mu }\varphi
)^{2}+\sum_{n=0}^{\infty }\!\!\!\!\!\!\left. \phantom{{a\over a}}\right.
^{\prime }\hbar ^{2n}[\partial ^{2n}]V_{n}(\varphi ,\hbar ).  \label{expa3}
\end{equation}
Indeed, formula (\ref{a1}) is corrected replacing $4L$ with $3L$, so 
\[
\omega (G)=L+2\sum_{n}nv_{n}-2V+2. 
\]
Moreover from (\ref{expa3}) the effective number of loops satisfies 
\[
L_{e}\geq L+2\sum_{n}nv_{n}, 
\]
whence 
\[
\omega (G)\leq L_{e}-2V+2. 
\]
Finally $V\geq 2$ implies 
\begin{equation}
\omega (G)\leq L_{e}-2.  \label{oi}
\end{equation}
The inequality (\ref{oi}) is the right inequality to prove that the
structure (\ref{expa3}) is preserved by renormalization, proceeding as
before. Only even $L$s have non-trivial logarithmic divergences in odd
dimensions, so $L_{e}$ is even. After appropriate field redefinitions the
irreducible counterterms of order $\hbar ^{L_{e}}$ have the form 
\[
\hbar ^{L_{e}}\sum_{k=0}^{L_{e}/2-1}\!\!\!\!\!\!\left. \phantom{{a\over a}}%
\right. ^{\prime }\frac{1}{\varepsilon ^{p(k)}}[\partial ^{2k}]\widetilde{H}%
_{L_{e},k}(\varphi ,\varepsilon ), 
\]
with $p(k)\leq L_{e}/2-k$ and are renormalized with the replacements 
\[
V_{n}(\varphi ,\hbar ,\varepsilon )\rightarrow V_{n}(\varphi ,\hbar
,\varepsilon )+\frac{\hbar ^{L_{e}-2n}}{\varepsilon ^{p(n)}}\widetilde{H}%
_{L_{e},n}(\varphi ,\varepsilon ),\qquad \text{for }n=0\text{ and }%
1<n<L_{e}/2, 
\]
thus preserving (\ref{expa3}). Finally, (\ref{imprexpa}) is replaced with 
\[
\mathcal{L}=\frac{M_{P}}{2}(\partial _{\mu }\chi )^{2}+\sum_{n=0}^{\infty
}\!\!\!\!\!\!\left. \phantom{{a\over a}}\right. ^{\prime }\hbar ^{2n}\alpha
^{2}M_{P}^{3-2n}[\partial ^{2n}]W_{n}(\hbar \alpha ,\chi ),\qquad
W_{n}(\hbar \alpha ,\chi )=\sum_{k=0}^{\infty }(\hbar \alpha
)^{2k}W_{n,k}(\chi ), 
\]
with $\chi =\varphi /M_{P}^{1/2}$.

Similar conclusions hold for fermionic theories of the form 
\begin{equation}
\mathcal{L}_{D=2}=\overline{\psi }\partial \!\!\!\slash \psi +V(\overline{%
\psi },\psi ,\lambda )  \label{fagra}
\end{equation}
in two spacetime dimensions. If the number of fermions is finite, a
non-derivative potential $V(\overline{\psi },\psi ,\lambda )$ necessarily
contains a finite number of terms, and so it is polynomial. However, the
construction of this section applies also when the number of fermions is
infinite (which is useful to study the large $N$ limit, for example), in
which case $V(\overline{\psi },\psi ,\lambda )$ is non-polynomial. The
complete renormalizable lagrangian has the form 
\begin{equation}
\mathcal{L}_{R}=\overline{\psi }\partial \!\!\!\slash \psi
+\sum_{n=0}^{\infty }\!\!\!\!\!\!\left. \phantom{{a\over a}}\right. ^{\prime
}\hbar ^{n}[\partial ^{n}]V_{n}(\overline{\psi },\psi ,\lambda ).
\label{fexpa}
\end{equation}

Coupled models with simplified renormalization structures exist also. An
example is the scalar-fermion theory 
\begin{equation}
\mathcal{L}=\frac{1}{2}(\partial _{\mu }\varphi )^{2}+\overline{\psi }%
\partial \!\!\!\slash \psi +\sum_{n,p=0}^{\infty }\!\!\!\!\!\!\left. %
\phantom{{a\over a}}\right. ^{\prime }\hbar ^{(n+p)/2}[\partial ^{n}](%
\overline{\psi }\psi )^{p}V_{np}(\varphi ,\hbar )  \label{ex}
\end{equation}
in four dimensions. Observe that the Yukawa interactions has a factor $\hbar
^{1/2}$. These developments and the study of applications to physics beyond
the Standard Model are left to a separate investigation.

\section{Perturbative expansion for the quantum action}

\setcounter{equation}{0}

In this section I discuss expansions compatible with the renormalization
structure (\ref{expa}),(\ref{imprexpa}). In a non-renormalizable theory it
is generically necessary to perform a double expansion: the loop expansion
and the expansion in powers $(E/M_{P})^{n}$ of the energy $E$, where $M_{P}$
is a reference scale for the irrelevant interactions. The expansion in
powers of the energy is an expansion in the dimensionality of the irrelevant
operators, which means in general an expansion in powers of the fields and
their derivatives. The loop expansion can be traded for the expansion in
powers of the marginal couplings, denoted collectively with $\alpha $, which
is ``orthogonal'' to the expansion in powers of the energy. Summarizing, the
ordinary expansion for a non-renormalizable theory is in powers of $\alpha $
and $E/M_{P}$, so it is necessary to assume that 
\begin{equation}
\alpha \ll 1,\qquad E\ll M_{P},  \label{normal}
\end{equation}
and $\varphi \sim E$, $\partial \varphi \sim E^{2}$, etc. An expansion of
this type applies also to the theories studied here, once the functions $%
V_{n}$ are expanded in powers of $\varphi $. However, when the $V_{n}$s are
treated non-perturbatively, new kinds of expansions apply.

Using the background field method \cite{backfmeth} the Feynman diagrams are
well defined when the action is non-polynomial or non-analytic in the
fields, and it is not expanded in powers of the fields, but only in their
derivatives. Indeed, expanding a generic function $V(\varphi )$ around a
background $\varphi _{0}$, $\varphi =\varphi _{0}+h$, 
\[
V(\varphi )=V(\varphi _{0})+hV^{\prime }(\varphi _{0})+\frac{1}{2}%
h^{2}V^{\prime \prime }(\varphi _{0})+\cdots , 
\]
each vertex has an integer number of $h$-legs, which are the internal legs
of the diagrams. The external ``legs'' need not be an integer number. They
are the functions $V^{\prime }(\varphi _{0}),V^{\prime \prime }(\varphi
_{0})\ldots $ When $V(\varphi )$ is not analytic the vertices are always
infinitely many, and contain also negative powers of $\varphi $. Therefore
the natural framework to study non-analytic theories is non-renormalizable
quantum field theory.

The renormalization structure (\ref{expa}),(\ref{imprexpa}) can be used in
two contexts: $i$) in the realm of effective field theory, where the
functions $V_{n}$ depend on infinitely many independent couplings; $ii$) in
the realm of fundamental field theory, where the number of independent
couplings can be consistently reduced. In the next section I prove that in
case $ii$) (\ref{imprexpa}) is replaced by 
\begin{equation}
\mathcal{L}=\frac{M_{P\text{eff}}^{2}}{2}(\partial _{\mu }\chi
)^{2}+\sum_{n=0}^{\infty }\!\!\!\!\!\!\left. \phantom{{a\over a}}\right.
^{\prime }\hbar ^{n}\alpha M_{P\text{eff}}^{4-2n}[\partial ^{2n}]W_{n}(\hbar
\alpha ,\chi ),  \label{vnform}
\end{equation}
where $M_{P\text{eff}}$ is a ``dressed'' Planck mass and now $\chi =\varphi
/M_{P\text{eff}}$.

I now prove that in both cases $i$) and $ii$) there exists a meaningful
perturbative expansion such that the physical quantities can be calculated
algorithmically with a finite number of steps at each order. I work with the
case $ii$) for definiteness, case $i$) being easily obtained from $ii$) with
the replacements $M_{P\text{eff}}\rightarrow M_{P}$, $\chi \rightarrow
\varphi /M_{P}$. For the time being I study a theory that does not contain
relevant parameters, so in particular there are no masses.

Consider the contributions $\Delta \Gamma _{G}=\int \delta \gamma _{G}$ to
the quantum action $\Gamma =\int \gamma $ due to the renormalized diagrams $%
G $ of order $L_{e}$ in $\hbar $, with $V$ vertices and $L$ loops. Since
there are no masses, the dependence of $\delta \gamma _{G}$ on the overall
energy $E$ is of the form $\delta \gamma _{G}(E,\mu )=$ $E^{\omega
(G)}\delta \widetilde{\gamma }_{G}(E/\mu )$, where $\mu $ is the subtraction
point. The $\mu $-dependence of $\delta \widetilde{\gamma }_{G}(E/\mu )$ is
only logarithmic, the maximal power of $\ln \mu $ being bounded by $L_{e}$.
Moreover, because of (\ref{vnform}), the power of $\alpha $ multiplying the
diagram is given by (\ref{obolo}). Collecting these pieces of information,
the $\alpha $-$E$-$M_{P\text{eff}}^{{}}$ dependence of $\delta \gamma _{G}$
is 
\begin{equation}
\delta \gamma _{G}=M_{P\text{eff}}^{4}\alpha ^{V+\Delta }(M_{P\text{eff}%
}^{{}}/E)^{-\omega (G)}\sum_{k\leq L_{e}}c_{k}\ln ^{\kappa }(E/\mu ),
\label{deltag}
\end{equation}
where $c_{k}$ are coefficients that depend on the dimensionless ratios among
the external momenta. The logarithms are reabsorbed into the running
couplings, including $\alpha $ and $M_{P\text{eff}}^{{}}$, and into the
running field. It is convenient to choose the subtraction point $\mu $ in
the same range as the energy $E$ (see below), because then the logarithms
are of order one. For simplicity, I keep the symbols $\alpha $ and $M_{P%
\text{eff}}^{{}}$ for the running constants. Using (\ref{degree}), we have 
\[
\alpha ^{V+\Delta }\left( \frac{M_{P\text{eff}}^{{}}}{E}\right) ^{-\omega
(G)}=\left( \frac{E}{M_{P\text{eff}}^{{}}}\right) ^{2(L_{e}+1)}\left( \frac{%
\alpha M_{P\text{eff}}^{2}}{E^{2}}\right) ^{V+\Delta }. 
\]
Now, the number of diagrams with given $L_{e}$ and $V$ is finite. Indeed, (%
\ref{usti}) implies $w_{n,k}=0$ for $n+k$ sufficiently large, so finitely
many types of vertices can be used to build the diagrams. Moreover, $%
w_{n,k}\leq V$ and $I=L+V-1\leq L_{e}+V-1$, so the number of vertices of
each type and the number of internal legs are finite. Since $0\leq \Delta
\leq L_{e}$, also the number of diagrams with given $L_{e}$ and $V+\Delta $
is finite. Thus if $E\ll M_{P\text{eff}}^{{}}$ and $\alpha M_{P\text{eff}%
}^{2}\ll E^{2}$ every quantity is computable with a finite number of steps
at each order of the expansion, both in the renormalization structure (\ref
{expa})\ and in the quantum action $\Gamma $.

Summarizing, in the absence of a mass gap, the expansion, at the level of
counterterms and in the quantum action $\Gamma $, is meaningful for energies
contained in the range 
\begin{equation}
\alpha M_{P\text{eff}}^{2}\ll E^{2}\ll M_{P\text{eff}}^{2},  \label{aret}
\end{equation}
while the field $\varphi $ is arbitrary, in particular it can be of order $%
M_{P\text{eff}}$. The marginal couplings $\alpha $ have to be small, so the
renormalizable subsector is weakly coupled. Observe that (\ref{aret})\
implies (\ref{normal}) (with $M_{P}\rightarrow M_{P\text{eff}}$), but not
viceversa.

In the presence of a mass gap $m$ some cases need to be distinguished. If $%
m^{2}\lesssim \alpha M_{P\text{eff}}^{2}$ then the mass gap can in practice
be neglected and everything works as before. In this case it is meaningful
to expand in powers of $m/E$. If $\alpha M_{P\text{eff}}^{2}\ll m^{2}\ll M_{P%
\text{eff}}^{2}$ it is meaningful to consider energies up to $E^{2}\ll M_{P%
\text{eff}}^{2}$. For energies $E^{2}\ll m^{2}$ it is possible to consider
the expansion in $E/m$ and the behavior of $\delta \gamma _{G}$ is certainly
bounded by (\ref{deltag}) with $E\rightarrow m$. Finally, when there is no
mass gap, but some particles are massive, it is sufficient to assume $\alpha
M_{P\text{eff}}^{2}\ll E^{2}\ll M_{P\text{eff}}^{2}$, $\alpha M_{P\text{eff}%
}^{2}\ll m^{2}\ll M_{P\text{eff}}^{2}$.

\section{Reduction of couplings}

\setcounter{equation}{0}

The rationalized renormalization structure (\ref{expa}),(\ref{imprexpa}) is
a remarkable restriction on non-re\-nor\-ma\-li\-zable theories, but still
contains infinitely many independent couplings. In this section I\ derive
general conditions to renormalize the divergences with a reduced, eventually
finite, number of independent couplings. It is sufficient to focus the
attention on the logarithmic divergences, setting the power-like divergences
aside. Indeed, the power-like divergences are RG invariant and can be
unambiguously subtracted away just as they come, without introducing new
independent couplings. The logarithmic divergences can be studied at the
level of the renormalization group, because the logarithms of the
subtraction point $\mu $ are in one-to-one correspondence with the
logarithms of the cut-off $\Lambda $.

The arguments of this section apply to the theories of the form (\ref{expa}%
), but immediate generalizations work also with the most general
non-renormalizable theory (see \cite{infredfin} for details). For
definiteness, I work in even $d$ spacetime dimensions. The generalization to
odd dimensions is straightforward, keeping in mind that, since the diagrams
with an odd number of loops have no logarithmic divergence, in the existence
conditions derived below the one-loop coefficients of the beta functions and
anomalous dimensions are replaced by two-loop coefficients, and so on. For
the moment I\ set $\hbar =1$.

Consider a strictly renormalizable theory $\mathcal{R}$ with marginal
couplings $\alpha $ and no relevant coupling. Let \textrm{O}$_{\lambda }$
denote a basis of ``essential'', local, symmetric, scalar, canonically
irrelevant operators constructed with the fields of $\mathcal{R}$ and their
derivatives. The essential operators are the equivalence classes of
operators that differ by total derivatives and terms proportional to the
field equations \cite{wein}. Total derivatives are trivial in perturbation
theory, while terms proportional to the field equations can be renormalized
away by means of field redefinitions, so they do not affect the beta
functions of the physical couplings. Finally, the operators \textrm{O}$%
_{\lambda }$ are Lorentz scalars and have to be ``symmetric'', that is to
say invariant under the non-anomalous symmetries of the theory, up to total
derivatives.

I assume that each \textrm{O}$_{\lambda }$ has a definite canonical
dimensionality $d_{\lambda }$ in units of mass. The irrelevant terms can be
ordered according to their ``level'' $\ell =d_{\lambda }-d$. It is
understood that in general each level contains several (but anyway finitely
many) operators, which can mix under renormalization. For the moment I do
not distinguish operators of the same level. I collectively denote the
irrelevant couplings of level $n$ with $\lambda _{n}$. On dimensional
grounds, the general form of the beta function $\beta _{n}$ is (see also 
\cite{pap2,pap3}) 
\begin{equation}
\beta _{n}=\lambda _{n}\gamma _{n}(\alpha )+\delta _{n},  \label{betacar}
\end{equation}
where $\delta _{n}$ depends only on $\lambda _{p}$ with $p<n$ and $\alpha $,
while $\gamma _{n}$ is the anomalous dimension of \textrm{O}$_{\lambda _{n}}$%
, calculated in the undeformed theory $\mathcal{R}$. Observe that a simple
structure such as (\ref{betacar}) follows from the assumption that no
relevant parameters are around. Define the marginal coupling $\alpha $ so
that the lowest $\alpha $-orders read 
\begin{eqnarray}
\beta _{\alpha } &=&\alpha ^{2}\beta _{0}^{(1)}+\mathcal{O}(\alpha
^{3}),\qquad \beta _{\ell }=\lambda _{\ell }\left( \alpha \gamma _{\ell
}^{(1)}+\mathcal{O}(\alpha ^{2})\right) ,  \label{tz} \\
\beta _{n} &=&\lambda _{n}\left( \alpha \gamma _{n}^{(1)}+\mathcal{O}(\alpha
^{2})\right) +\delta _{n}(\lambda ,\alpha ),\qquad n>\ell .  \label{tz2}
\end{eqnarray}

Consider an irrelevant deformation with ``head'' $\lambda _{\ell }$ of level 
$\ell $. The head of the deformation is the irrelevant term with the lowest
dimensionality. The ``queue'' of the deformation is made of the irrelevant
terms that have dimensionalities greater than $\ell $, whose couplings are
not independent, but functions of $\alpha $ and $\lambda _{\ell }$: 
\begin{equation}
\mathcal{L}_{\mathrm{cl}}[\varphi ]=\mathcal{L}_{\mathcal{R}}[\varphi
,\alpha ]+\lambda _{\ell }\mathcal{O}_{\ell }(\varphi )+\sum_{n=2}^{\infty
}\lambda _{n\ell }(\lambda _{\ell },\alpha )\mathcal{O}_{n\ell }(\varphi ).
\label{beh2}
\end{equation}
By dimensional arguments, the queue of the deformation is made only of terms
whose levels are integer multiples of $\ell $. All other $\lambda _{n}$s are
set to zero.

The goal is to find a ``reduction of couplings'', namely a set of functions $%
\lambda _{n\ell }(\lambda _{\ell },\alpha )$, $n>1$, such that the theory is
renormalized by means of field redefinitions plus renormalization constants
for $\alpha $ and $\lambda _{\ell }$. Dimensional arguments ensure that the
functions $\lambda _{n\ell }(\lambda _{\ell },\alpha )$ have the form 
\begin{equation}
\lambda _{n\ell }=f_{n}(\alpha )\lambda _{\ell }^{n}  \label{anz}
\end{equation}
and consequently the form of the beta functions (\ref{betacar}) is 
\begin{equation}
\beta _{n\ell }=\lambda _{\ell }^{n}\left[ f_{n}(\alpha )\gamma _{n\ell
}(\alpha )+\delta _{n}(f,\alpha )\right] ,  \label{gret}
\end{equation}
where $\delta _{n}(f,\alpha )$ depends only on $f_{k}$ with $k<n$. Formulas (%
\ref{anz}) and (\ref{gret}) hold also for $n=1$, with $f_{1}=1$ and $\delta
_{1}=0$. Differentiating the functions (\ref{anz}) with respect to the
dynamical scale $\mu $ and using the definitions of beta functions, the RG
consistency equations 
\begin{equation}
f_{n}^{\prime }(\alpha )\beta _{\alpha }-f_{n}(\alpha )\left[ \gamma _{n\ell
}(\alpha )-n\gamma _{\ell }(\alpha )\right] =\delta _{n}(f,\alpha )
\label{rg}
\end{equation}
are obtained. Now I show that these equations are necessary and sufficient
conditions to renormalize the logarithmic divergences of the theory by means
of renormalization constants just for $\lambda _{\ell }$ and $\alpha $, plus
field redefinitions.

Write the bare couplings $\lambda _{n\ell }(\Lambda )$ and $\alpha (\Lambda
) $ in terms of their renormalization constants $Z_{n\ell }$ and $Z_{\alpha
} $ in the minimal subtraction scheme, where $\lambda _{n\ell }$ and $\alpha 
$ are the renormalized couplings at the subtraction point $\mu $: 
\[
\lambda _{n\ell }(\Lambda )=\lambda _{n\ell }Z_{n\ell }(\lambda ,\alpha ,\ln
\Lambda /\mu ),\qquad \alpha (\Lambda )=\alpha Z_{\alpha }(\alpha ,\ln
\Lambda /\mu ). 
\]
Now, assume that $\lambda _{n\ell }$ satisfy (\ref{anz}) and (\ref{rg}). The
RG\ consistency conditions (\ref{rg}) imply that the reduction relations
have the same form at every energy scale, in particular $\mu $ and $\Lambda $%
. Consequently, 
\begin{equation}
\lambda _{n\ell }Z_{n\ell }=\lambda _{n\ell }(\Lambda )=f_{n}(\alpha
(\Lambda ))\lambda _{\ell }^{n}(\Lambda )=\lambda _{\ell }^{n}Z_{\ell
}^{n}f_{n}(\alpha Z_{\alpha }),\qquad n>1.  \label{areno}
\end{equation}
Thus the couplings $\lambda _{n\ell }$, $n>1$, can be renormalized just
attaching renormalization constants to $\lambda _{\ell }$ and $\alpha $.

However, no true reduction of couplings is achieved simply solving the RG
consistency conditions (\ref{rg}). Indeed, (\ref{rg}) are differential
equations for the unknown functions $f_{n}(\alpha )$, $n>1$. The solutions
depend on arbitrary constants $\xi $. From the point of view of the
renormalization, the arbitrary constants $\xi $ are finite parameters,
namely $Z_{\xi }\equiv 1$. The equations (\ref{rg}) and the arguments
leading to (\ref{areno}) are simply a rearrangement of renormalization, with
no true gain, because the number of renormalization constants is reduced at
the price of introducing new constants $\xi _{n}$. To remove the $\xi $%
-ambiguity contained in the solutions of (\ref{rg}) and achieve a true
reduction of couplings, extra assumptions have to be made.

A similar problem appears in the context of renormalizable theories, where
Zimmermann proposed to eliminate the $\xi $-arbitrariness requiring that the
reduction relations be analytic \cite{zimme}. However, in the realm of
non-renormalizable theories analyticity is a too restrictive requirement:
negative powers of the marginal coupling $\alpha $ appear \cite{pap2,pap3}
and are reabsorbed into the effective Planck mass $M_{P\text{eff}}$. Thus
the correct requirement is meromorphy.

For example, in the beta function $\beta _{2\ell }$ the coupling $\lambda
_{2\ell }$ is multiplied by $\gamma _{2\ell }=\mathcal{O}(\alpha )$, thus $%
f_{2}(\alpha )$ has the form 
\begin{equation}
f_{2}(\alpha )=\frac{1}{\alpha }\sum_{k=0}^{\infty }d_{2,k}\alpha ^{k}.
\label{ansa}
\end{equation}
Inserting the ansatz (\ref{ansa}) into (\ref{rg}), using (\ref{tz}) and
solving for $d_{2,k}$ recursively in $k$, it is easy to check that the
coefficients $d_{2,k}$ have the form 
\begin{equation}
d_{2,k}=\frac{P_{2,k}}{\prod_{j=1}^{k+1}\left( \gamma _{2\ell
}^{(1)}-2\gamma _{\ell }^{(1)}+(2-j)\beta _{0}^{(1)}\right) }.  \label{dena}
\end{equation}
The numerator $P_{2,k}$ depends polynomially on the coefficients of the beta
functions $\beta _{n\ell },\beta _{\alpha }$ and in general does not vanish
when one of the factors appearing in the denominator vanishes. Thus,
assuming that $\beta _{0}^{(1)}\neq 0$, the meromorphic solution (\ref{ansa}%
) is meaningful, and unique, when the quantity 
\begin{equation}
r_{2,\ell }\equiv \frac{\gamma _{2\ell }^{(1)}-2\gamma _{\ell }^{(1)}}{\beta
_{0}^{(1)}}+1  \label{into}
\end{equation}
is not a natural number.

Now I prove that, if $\beta _{0}^{(1)}\neq 0$ and 
\begin{equation}
r_{n,\ell }\equiv \frac{\gamma _{n\ell }^{(1)}-n\gamma _{\ell }^{(1)}}{\beta
_{0}^{(1)}}+n-1\notin \mathbb{N},\qquad n>1,  \label{conditions}
\end{equation}
there exists a unique meromorphic reduction such that the couplings $\lambda
_{k\ell }$ behave at worst as \cite{pap2,pap3} 
\begin{equation}
\lambda _{k\ell }\sim c_{k}\frac{\lambda _{\ell }^{k}}{\alpha ^{k-1}},
\label{behave}
\end{equation}
for small $\alpha $, where $c_{k}$ are constants. This result is proved by
induction. It is certainly true for $k=1$ and follows from (\ref{ansa}) for $%
k=2$. Assume that it is true for $k<n$. Since $\delta _{n}$ depends on the $%
\lambda $s with lower levels, we have, by dimensional analysis, 
\[
\delta _{n\ell }\sim \sum_{\{n_{k}\}}\prod_{k<n}\lambda _{k\ell
}{}^{n_{k}}\left( 1+\mathcal{O}(\alpha )\right) , 
\]
where the sum is made over the sets $\{n_{k}\}$ of non-negative integers $%
n_{k}$ such that $\sum_{k<n}kn_{k}=n$. Moreover, $m\equiv
\sum_{k<n}n_{k}\geq 2$, since $\delta _{n}$ is at least quadratic. Therefore 
\begin{equation}
\delta _{n\ell }\sim \lambda _{\ell }^{n}\sum_{\{n_{k}\}}\prod_{k<n}\left( 
\frac{1}{\alpha ^{k-1}}\right) ^{n_{k}}=\lambda _{\ell }^{n}\sum_{\{n_{k}\}}%
\frac{1}{\alpha ^{n-m}}\lesssim \lambda _{\ell }^{n}\frac{1}{\alpha ^{n-2}}.
\label{esti}
\end{equation}
The general form of the meromorphic reduction relation is 
\begin{equation}
f_{n}(\alpha )=\frac{1}{\alpha ^{p}}\sum_{k=0}^{\infty }d_{n,k}\alpha ^{k}.
\label{beh1}
\end{equation}
Inserting this ansatz into (\ref{rg}), using (\ref{beh1}) and solving for $%
d_{n,k}$ recursively in $k$ it is immediate to find that $p=n-1$ and the
coefficients $d_{n,k}$ have expressions 
\begin{equation}
d_{n,k}=\frac{P_{n,k}}{\prod_{j=1}^{k+1}\left( \gamma _{n\ell
}^{(1)}-n\gamma _{\ell }^{(1)}+(n-j)\beta _{0}^{(1)}\right) },  \label{dnk}
\end{equation}
where $P_{n,k}$ depends polynomially on the coefficients of the beta
functions and on $d_{m,k}$, $m<n$. In general the numerator $P_{n,k}$ does
not vanish when the denominator vanishes. Thus (\ref{behave}) is inductively
proved for arbitrary $n$.

In conclusion, when the invertibility conditions (\ref{conditions}) are
fulfilled, there exists a unique meromorphic reduction with the behavior (%
\ref{behave}).

The quantities $r_{n,\ell }$ depend only on one-loop coefficients, yet the
conditions (\ref{conditions}) determine the existence of the meromorphic
reduction to all orders. Moreover, the $r_{n,\ell }$s are just rational
numbers and it is not unfrequent that they coincide with natural numbers for
some $n$s, so sometimes the invertibility conditions are violated.

\bigskip

\textbf{Violations of the invertibility conditions}. Suppose that $r_{%
\overline{n},\ell }$ is a natural number $\overline{k}$ for some $\overline{n%
}$. Then, (\ref{dnk}) shows that the reduction fails at the $\overline{k}$th
order in $\alpha $. This problem is avoided introducing a new independent
parameter $\overline{\lambda }_{\overline{n}\ell }$ at order $\overline{k}$
in front of $\mathcal{O}_{\overline{n}\ell }$, writing 
\begin{equation}
\lambda _{\overline{n}\ell }=\frac{1}{\alpha ^{\overline{n}-1}}\left[
\lambda _{\ell }^{\overline{n}}\sum_{j=0}^{\overline{k}-1}d_{\overline{n}%
,j}\alpha ^{j}+\alpha ^{\overline{k}}\overline{\lambda }_{\overline{n}\ell
}\right] ,  \label{tre}
\end{equation}
where $d_{\overline{n},j}$, $j<\overline{k}$ are calculated as above. The
beta function of $\overline{\lambda }_{\overline{n}\ell }$ has the form 
\[
\overline{\beta }_{\overline{n}\ell }=\overline{\gamma }_{\overline{n}\ell
}\left( \alpha \right) \overline{\lambda }_{\overline{n}\ell }+\overline{%
\delta }_{\overline{n}\ell }(\alpha ,\lambda _{m<\overline{n}}),\qquad 
\overline{\gamma }_{\overline{n}\ell }\left( \alpha \right) =\overline{n}%
\gamma _{\ell }^{(1)}\alpha +\mathcal{O}(\alpha ^{2}),\qquad \overline{%
\delta }_{\overline{n}\ell }=\lambda _{\ell }^{\overline{n}}\mathcal{O}%
(\alpha ). 
\]
The one-loop coefficient of $\overline{\gamma }_{\overline{n}\ell }$ is
derived from $r_{\overline{n},\ell }=\overline{k}$.

The new parameter $\overline{\lambda }_{\overline{n}\ell }$ modifies the
reduction relations also for $n>\overline{n}$. Taking into account that $%
\overline{\lambda }_{\overline{n}\ell }$ contributes only from order $%
\overline{k}$, the modified reduction relations read 
\begin{equation}
\lambda _{n\ell }=\frac{1}{\alpha ^{n-1}}\sum_{q=0}^{[n/\overline{n}]}\alpha
^{\overline{k}q}a_{n\ell }^{(q)}(\alpha )\lambda _{\ell }^{n-\overline{n}q}%
\overline{\lambda }_{\overline{n}\ell }^{q},\qquad n>\overline{n},
\label{nego}
\end{equation}
where $[$ $]$ denotes the integral part and the coefficients $a_{n\ell
}^{(q)}$ are power series in $\alpha $. Inserting (\ref{nego}) in (\ref{gret}%
) the coefficients $a_{n\ell }^{(q)}$ are worked out recursively, from $q=[n/%
\overline{n}]$ to $q=0$, term-by-term in the $\alpha $-expansion. The
existence conditions for $a_{nm}^{(q)}$, namely 
\begin{equation}
r_{n,\ell ,q}=r_{n,\ell }-\overline{k}q\notin \mathbb{N},  \label{klq}
\end{equation}
do not add further restrictions, because they are already contained in (\ref
{conditions}).

When another invertibility condition (\ref{conditions}), $n>\overline{n}$,
is violated, the story repeats. A new parameter $\overline{\lambda }_{n\ell
} $ is introduced at order $\alpha ^{r_{n,\ell }}$. If several conditions (%
\ref{klq}), for different values of $q$, are violated at the same time, all
relevant monomials of (\ref{nego}) are reabsorbed into the same new
parameter $\overline{\lambda }_{n\ell }$.

\bigskip

\textbf{Effects of the renormalization mixing.} Consider the renormalization
mixing, calculated in the undeformed theory $\mathcal{R}$, among operators
with the same dimensionality $n\ell $ in units of mass, $n\geq 1$.
Distinguish the mixing operators with indices $I,J,\ldots $. If $\ell $ is
the level of the deformation, denote the inequivalent operators of level $%
\ell $ with O$_{\ell }^{I}$, the coefficient-matrix of their lowest-order
anomalous dimensions with $\gamma _{\ell }^{(1)\ IJ}$, an eigenvalue of $%
\gamma _{\ell }^{(1)\ IJ}$ with $\gamma _{\ell }^{(1)}$ and the
corresponding eigenvector with $d_{0}^{I}$. Denote the operators of the
queue with O$_{n\ell }^{I}$, $n>1$, and their couplings with $\lambda
_{n\ell }^{I}$. The beta functions read 
\[
\beta _{n\ell }^{I}=\sum_{J}\gamma _{n\ell }^{IJ}(\alpha )\lambda _{n\ell
}^{J}+\delta _{n\ell }^{I}, 
\]
where $\delta _{n\ell }^{I}$ depends only on $\lambda _{m\ell }^{I}$ with $%
m<n$ and $\alpha $. Introduce an auxiliary coupling $\lambda _{\ell }$ of
level $\ell $, with beta function $\beta _{\lambda _{\ell }}=\gamma _{\ell
}^{(1)}\alpha \lambda _{\ell }$. The beta function of $\lambda _{\ell }$ can
be chosen to be one-loop exact with an approprite scheme choice (any other
choice being equivalent to a redefinition $\lambda _{\ell }\rightarrow
h(\alpha )\lambda _{\ell }$, with $h(\alpha )$ analytic in $\alpha $, $%
h(0)=1 $). The reduction relations have the form 
\[
\lambda _{n\ell }^{I}=f_{n}^{I}(\alpha )\lambda _{\ell }^{n},\qquad n\geq 1, 
\]
where $f_{n}^{I}(\alpha )=\mathcal{O}(\alpha ^{1-n})$. If $k$ is a natural
number, the existence conditions are that the matrices 
\begin{equation}
r_{n,k,\ell }^{IJ}=\gamma _{n\ell }^{(1)\ IJ}-n\gamma _{\ell }^{(1)}\delta
^{IJ}+(n-1-k)\beta _{0}^{(1)}\delta ^{IJ}  \label{matrix}
\end{equation}
be invertible for $n>1$, $k\geq 0$ and for $n=1$, $k>0$. If the
invertibility conditions are fulfilled, the solution is uniquely determined
in terms of $d_{0}^{I}$. The head of the deformation is $\sum_{I}$O$_{\ell
}^{I}\lambda _{\ell }^{I}$.

A lowest-order non-trivial renormalization mixing makes the existence
conditions much easier to fulfill. Indeed, the entries of the matrices $%
\gamma _{n\ell }^{(1)\ IJ}$ are rational numbers divided by $\pi ^{d/2}$.
Non-trivial (i.e. non-triangular) matrices with rational entries have in
general irrational, or possibly complex, eigenvalues. The invertibility of $%
r_{n,k,\ell }^{IJ}$ requires that certain linear combinations of generically
irrational or complex numbers do not coincide with natural numbers. The
violations of this requirement are much rarer than the violations of (\ref
{conditions}).

In the theories considered here, which have renormalization structures such
as (\ref{master}), the matrices $r_{n,k,\ell }^{IJ}$ are block-triangular.
More precisely, at the lowest order the renormalization mixing calculated in
the undeformed theory $\mathcal{R}$ can be non-triangular only among
operators that have the same number of derivatives. Write 
\begin{equation}
\mathrm{O}_{p,q}=[\partial ^{2p}]\varphi ^{q}\text{,}  \label{opa}
\end{equation}
where, as usual, derivatives are distributed and contracted in all possible
ways, up to total derivatives and terms factorizing $\Box \varphi $. Each $%
\mathrm{O}_{p,q}$ is a set of operators $\mathrm{O}_{p,q}^{I}$, where $I$
labels the independent contractions of derivatives. Now I\ prove that at the
lowest-order the matrix $\gamma _{p,q|p^{\prime },q^{\prime }}^{I|J}$ of
anomalous dimensions (calculated in the undeformed theory $\mathcal{R}$,
that is to say the four-dimensional $\varphi ^{4}$ theory) vanishes for $%
p<p^{\prime }$. To simplify the notation, call $\gamma _{pp^{\prime }}^{(1)}$
the lowest-order coefficients of this matrix and collectively denote with $%
\mathrm{O}_{p}=[\partial ^{2p}]F_{p}(\varphi )$ the set of operators $%
\mathrm{O}_{p,q}^{I}$ that have $2p$ derivatives. Study the lowest-order
counterterms of type $\mathrm{O}_{p}$ generated by the diagrams that contain
one insertion of $\mathrm{O}_{p^{\prime }}$. This amounts to set $v_{0}=1$, $%
v_{p^{\prime }}=1$, $v_{n}=0$ for $n\neq p^{\prime }$, $n\neq 0$ and $L\geq
1 $ in (\ref{obaoba}), so 
\[
p=\frac{1}{2}\omega (G)=p^{\prime }+L-1\geq p^{\prime }. 
\]
Thus $\gamma _{pp^{\prime }}^{(1)}=0$ for $p<p^{\prime }$, which proves the
claimed block-triangular structure of the matrix of lowest-order anomalous
dimensions.

Obviously, there is only one operator (\ref{opa}) with given $q$ and no
derivative, that is $\varphi ^{n\ell +4}$, so the first diagonal block of (%
\ref{matrix}) is a one-by-one matrix and coincides with $(r_{n,\ell
}-k)\beta _{0}^{(1)}$, $k\in \mathbb{N}$: a necessary condition for the
invertibility of the matrices (\ref{matrix}) is still $r_{n,\ell }\notin %
\mathbb{N}$. The eigenvector $d_{0}^{I}$ corresponds to the operator $%
\varphi ^{\ell +4}$, which is the head of the deformation, and $\gamma
_{\ell }^{(1)}$ is the coefficient of its one-loop anomalous dimension.
Operators (\ref{opa}) with $q=3$ are either relevant ($p=0$) or factorize a $%
\Box \varphi $. Operators with $p=1$ are trivial. There is a unique operator
with $p=2$ and given $q$, e.g. 
\begin{equation}
\varphi ^{q-4}\left[ (\partial _{\mu }\varphi )^{2}\right] ^{2},
\label{affo}
\end{equation}
which provides invertibility conditions similar to (\ref{conditions}).
Instead, starting from $p=3$ there are at least two inequivalent operators
for every $q$, e.g. for $p=3$%
\begin{equation}
\varphi ^{q-6}\left[ (\partial _{\mu }\varphi )^{2}\right] ^{3},\qquad
\varphi ^{q-4}(\partial _{\mu }\varphi )^{2}(\partial _{\nu }\partial _{\rho
}\varphi )^{2},  \label{affa}
\end{equation}
so the diagonal blocks with $p>2$ provide less restrictive invertibility
conditions. If some matrix $r_{n,k,\ell }^{IJ}$ exceptionally has a null
eigenvector, the new parameter is introduced with a procedure similar to the
one explained in formulas (\ref{tre}) and (\ref{nego}).

\bigskip

Summarizing, neither the extra parameters $\overline{\lambda }_{\overline{n}%
\ell }$ nor the renormalization mixing modify (\ref{conditions}). The
structure of the invertibility conditions is always (\ref{conditions}) or (%
\ref{matrix}).

In the next sections I apply these considerations to the theory (\ref{expa}%
), at one and two loops, and show that in common models (analytic
potentials) the existence conditions are violated in an infinity of cases,
namely there exist infinitely many $\overline{n}$s such that $r_{\overline{n}%
,\ell }=\overline{k}(\overline{n})\in \mathbb{N}$. The final theory contains
infinitely many independent couplings, but each new coupling is introduced
at an order $\overline{k}(\overline{n})$ in $\alpha $ that typically grows
with $\overline{n}$, so it is possible to make calculations up to high
orders using a relatively small number of couplings. I identify also the
cases where the existence conditions do not admit any violations.

\bigskip

\textbf{Role of the effective Planck mass.} Finally, I prove (\ref{vnform}).
Start from the renormalization structure (\ref{master}) in four dimensions.
Restore the $\hbar $-dependence and expand 
\begin{equation}
\hbar ^{p}V_{p}(\varphi ,\hbar )=\sum_{q\geq 4}\lambda _{p,q}\varphi ^{q}.
\label{bah}
\end{equation}
Consider an irrelevant deformation of level $\ell $. The non-vanishing
irrelevant couplings are such that $n_{p,q}\equiv (2p+q-4)/\ell $ is integer 
$\geq 1$. Decompose the beta functions in the usual form, $\beta
_{p,q}=\lambda _{p,q}\gamma _{p,q}(\hbar \alpha )+\delta _{p,q}(\lambda
_{p^{\prime },q^{\prime }},\alpha ,\hbar )$, ignoring for the moment the
renormalization-mixing, where $\delta _{p,q}$ depends only on the couplings $%
\lambda _{p^{\prime },q^{\prime }}$ with $n_{p^{\prime },q^{\prime
}}<n_{p,q} $.

Precisely, $\delta _{p,q}$ receives contributions form the diagrams $G$ such
that $\omega (G)/2=p$. If $v_{0}$ is the number of marginal vertices and $%
v_{p^{\prime },q^{\prime }}$ is the number of irrelevant vertices of type $%
p^{\prime },q^{\prime }$, $\delta _{p,q}$ has the structure 
\begin{equation}
\delta _{p,q}\sim \sum_{G}\hbar ^{L}\alpha ^{v_{0}}\prod_{n_{p^{\prime
},q^{\prime }}<n_{p,q}}\lambda _{p^{\prime },q^{\prime }}^{v_{p^{\prime
},q^{\prime }}},  \label{adelta}
\end{equation}
where $v_{p^{\prime },q^{\prime }}$ are such that $\sum_{n_{p^{\prime
},q^{\prime }}<n_{p,q}}v_{p^{\prime },q^{\prime }}n_{p^{\prime },q^{\prime
}}=n_{p,q}$, while $\sum_{n_{p^{\prime },q^{\prime }}<n_{p,q}}v_{p^{\prime
},q^{\prime }}=V-v_{0}$ is the total number of irrelevant vertices. The
condition $\omega (G)/2=p$ gives, using (\ref{obaoba}), $L+\sum_{n_{p^{%
\prime },q^{\prime }}<n_{p,q}}p^{\prime }v_{p^{\prime },q^{\prime }}-V+1=p$.

If the invertibility conditions are fulfilled, the solution to the reduction
equations is $\lambda _{p,q}=f_{p,q}(\alpha ,\hbar )\lambda _{\ell
}^{n_{p,q}}$, with 
\begin{equation}
f_{p,q}(\alpha ,\hbar )=\frac{\hbar ^{p}}{\alpha ^{n_{p,q}-1}}%
\sum_{k=0}^{\infty }(\hbar \alpha )^{k}d_{p,q,k},  \label{behave2}
\end{equation}
$d_{p,q,k}$ being uniquely determined numerical coefficients. Indeed,
assuming inductively that (\ref{behave2}) holds for $p,q$ such that $%
n_{p,q}<n_{\overline{p},\overline{q}}$, (\ref{adelta}) implies 
\[
\delta _{\overline{p},\overline{q}}\sim \sum_{G}\hbar ^{L}\alpha
^{v_{0}}\lambda _{\ell }^{n_{\overline{p},\overline{q}}}\prod_{n_{p,q}<n_{%
\overline{p},\overline{q}}}\left( \frac{\hbar ^{p}}{\alpha ^{n_{p,q}-1}}%
\right) ^{v_{p,q}}(1+\mathcal{O}(\hbar \alpha )), 
\]
the diagrams $G$ being such $\omega (G)/2=\overline{p}$. Therefore, 
\[
\delta _{\overline{p},\overline{q}}\sim \hbar ^{\overline{p}}\frac{\lambda
_{\ell }^{n_{\overline{p},\overline{q}}}}{\alpha ^{n_{\overline{p},\overline{%
q}}-1}}\sum_{G}(\hbar \alpha )^{V-1}(1+\mathcal{O}(\hbar \alpha )). 
\]
Recalling that $\lambda _{p,q}\sim \delta _{p,q}/(\hbar \alpha )$ and $V\geq
2$, the result (\ref{behave2}) follows for $\overline{p},\overline{q}$.

It is easy to show that the result holds also when the
renormalization-mixing is taken into account. This check is left to the
reader.

Finally, inserting (\ref{behave2}) into (\ref{bah}) and (\ref{expa}), the
complete lagrangian reads 
\begin{equation}
\mathcal{L}=\frac{1}{2}(\partial \varphi )^{2}+\sum_{p=0}^{\infty
}\!\!\!\!\!\!\left. \phantom{{a\over a}}\right. ^{\prime }\hbar
^{p}\sum_{n\geq 2p/\ell }\frac{\lambda _{\ell }^{n}}{\alpha ^{n-1}}%
w_{p,n}(\hbar \alpha )[\partial ^{2p}]\varphi ^{n\ell -2p+4},  \label{beh3}
\end{equation}
where $w_{n,p}(\hbar \alpha )$ are power series in $\hbar \alpha $. The
negative powers of $\alpha $ are reabsorbed into the effective Planck mass $%
M_{P\text{eff}}$. Defining 
\begin{equation}
\lambda _{\ell }=\frac{1}{M_{P}^{\ell }},\qquad M_{P\text{eff}}^{\ell
}=M_{P}^{\ell }\alpha ,\qquad \chi =\frac{\varphi }{M_{P\text{eff}}},\qquad
W_{n}(\alpha ,\chi ,\hbar )=\sum_{s\geq 2n/\ell }w_{n,s}(\hbar \alpha )\chi
^{s\ell -2n+4},  \label{oss}
\end{equation}
(\ref{beh3}) becomes (\ref{vnform}).

Due to (\ref{tre}) and (\ref{nego}), the introduction of new parameters $%
\overline{\lambda }_{\overline{n}\ell }$ does not change this structure: it
is sufficient to define $\overline{\lambda }_{\overline{n}\ell \text{eff}%
}=\alpha ^{-\overline{n}}\overline{\lambda }_{\overline{n}\ell }$ and take $%
\alpha $ small at fixed $M_{P\text{eff}}$ and $\overline{\lambda }_{%
\overline{n}\ell \text{eff}}$.

The divergences of the reduced theory (\ref{vnform}) are reabsorbed into
renormalization constants for the independent couplings, $\alpha $, $M_{P%
\text{eff}}$ and eventually $\overline{\lambda }_{\overline{n}\ell \text{eff}%
}$, plus field redefinitions. It is easy to show that the form of the field
redefinition is the same as (\ref{frefa}) with $M_{P}\rightarrow M_{P\text{%
eff}}$ and $\chi \rightarrow \varphi /M_{P\text{eff}}$. I leave the proof to
the reader.

\section{The $\varphi ^{4}+\varphi ^{6}$ theory in four dimensions}

\setcounter{equation}{0}

Now I\ study the theory 
\begin{equation}
\mathcal{L}_{\text{cl}}=\frac{1}{2}(\partial _{\mu }\varphi )^{2}+V_{\text{cl%
}}(\varphi ,\lambda ),\qquad V_{\text{cl}}(\varphi ,\lambda )=\frac{\lambda
_{0}}{4!}\varphi ^{4}+\frac{\lambda _{2}}{6!}\varphi ^{6}\text{,}
\label{claf}
\end{equation}
in four dimensions, using the tools developed in the previous sections. The
theory (\ref{claf}) is classically meaningful, but as a quantum field theory
it is not consistent, since the $\varphi ^{6}$ interaction is unstable under
renormalization. It is necessary to look for a completion of type (\ref{expa}%
) that is stable under renormalization, such that the divergences are
removed only with field redefinitions and renormalization constants for $%
\lambda _{0}$ and $\lambda _{2}$. In this section I study this completion
and one and two loops, non-perturbatively in $\varphi $.

\bigskip

\textbf{Structure of the lagrangian.} The bare lagrangian has generically
the form 
\begin{equation}
\mathcal{L}_{\mathrm{B}}=\frac{1}{2}(\partial _{\mu }\varphi _{\mathrm{B}%
})^{2}+\sum_{n=0}^{\infty }\!\!\!\!\!\!\left. \phantom{{a\over a}}\right.
^{\prime }[\partial ^{2n}]V_{\mathrm{B}n}(\varphi _{\mathrm{B}},\lambda _{%
\mathrm{B}},\varepsilon ).  \label{bare}
\end{equation}
The information (\ref{expa}) that for $n>1$, $2n$ derivatives of the fields
do not appear before the $n$th loop will be used later. The functions $V_{%
\mathrm{B}n}(\varphi _{\mathrm{B}},\lambda _{\mathrm{B}},\varepsilon )$ can
depend on $\varepsilon $, but only analytically, since the divergences are
reabsorbed inside $\varphi _{\mathrm{B}}$ and $\lambda _{\mathrm{B}}$, by
definition. The $\varepsilon $-dependence of (\ref{bare}) is important for
the reduction of couplings discussed below, since the $\varepsilon $-powers
of (\ref{bare}) can simplify divergences and give finite contributions.

Now I study the constraints on $\mathcal{L}_{\mathrm{B}}$ due to
renormalization stability, using the dimensional-regularization technique,
where $d=4-\varepsilon $ is the continued dimension. The dimensionalities of
objects are denoted by means of square brackets, so 
\[
\lbrack \varphi _{\mathrm{B}}]=1-\frac{\varepsilon }{2},\qquad [\lambda _{0%
\mathrm{B}}]=\varepsilon ,\qquad [\lambda _{2\mathrm{B}}]=-2+2\varepsilon . 
\]
With the three dimensionful quantities $\lambda _{0\mathrm{B}},\lambda _{2%
\mathrm{B}}$ and $\varphi _{\mathrm{B}}$ it is possible to construct two
dimensionless combinations 
\begin{equation}
\frac{\varphi _{\mathrm{B}}^{2}\lambda _{2\mathrm{B}}}{\lambda _{0\mathrm{B}}%
},\qquad \lambda _{0\mathrm{B}}^{2(\varepsilon -1)}\lambda _{2\mathrm{B}%
}^{-\varepsilon }.  \label{comba}
\end{equation}
One of them, however, does not depend on $\varphi _{\mathrm{B}}$ and has $%
\varepsilon $-dependent exponents, so it can generate logarithms and
fractional powers of the couplings in the limit $d\rightarrow 4$. Since the
reduction is meromorphic, the functions $V_{\mathrm{B}n}(\varphi _{\mathrm{B}%
},\lambda _{\mathrm{B}},\varepsilon )$ can depend only on the first quantity
of (\ref{comba}). Assume for simplicity that $\lambda _{0}$ and $\lambda
_{2} $ are positive. Defining 
\[
\chi _{\mathrm{B}}=\varphi _{\mathrm{B}}\sqrt{\frac{\lambda _{2\mathrm{B}}}{%
\lambda _{0\mathrm{B}}}} 
\]
and matching the dimensionalities at $\varepsilon \neq 0$, the functions $V_{%
\mathrm{B}n}$ have the form 
\begin{equation}
V_{\mathrm{B}n}(\varphi _{\mathrm{B}},\lambda _{\mathrm{B}},\varepsilon )=%
\frac{\lambda _{2\mathrm{B}}^{n-2}}{\lambda _{0\mathrm{B}}^{2n-3}}W_{\mathrm{%
B}n}(\chi _{\mathrm{B}},\varepsilon ).  \label{obla}
\end{equation}
Again, the functions $W_{\mathrm{B}n}(\chi _{\mathrm{B}},\varepsilon )$ can
depend on $\varepsilon $ only analytically, since the divergences are
reabsorbed inside $\chi _{\mathrm{B}},\lambda _{\mathrm{B}}$.

Now I\ show that the property (\ref{renexpa}) that for $n>1$, $2n$
derivatives of the fields do not appear in the renormalized lagrangian
before the $n$th loop is equivalent to $W_{\mathrm{B}n}=\mathcal{O}%
(\varepsilon ^{n})$. Write $W_{\mathrm{B}n}(\chi _{\mathrm{B}},\varepsilon
)=\varepsilon ^{n}\overline{W}_{\mathrm{B}n}(\chi _{\mathrm{B}},\varepsilon
) $, with $\overline{W}_{\mathrm{B}n}(\chi _{\mathrm{B}},\varepsilon )$
analytic in $\varepsilon $. The bare lagrangian (\ref{bare}) reads 
\begin{equation}
\mathcal{L}_{\mathrm{B}}=\frac{\lambda _{0\mathrm{B}}}{2\lambda _{2\mathrm{B}%
}}(\partial _{\mu }\chi _{\mathrm{B}})^{2}+\sum_{n=0}^{\infty
}\!\!\!\!\!\!\left. \phantom{{a\over a}}\right. ^{\prime }\varepsilon ^{n}%
\frac{\lambda _{2\mathrm{B}}^{n-2}}{\lambda _{0\mathrm{B}}^{2n-3}}[\partial
^{2n}]\overline{W}_{\mathrm{B}n}(\chi _{\mathrm{B}},\varepsilon ).
\label{bare2}
\end{equation}
At the tree level bare and renormalized quantities coincide, so (\ref{bare2}%
) corresponds to the tree lagrangian 
\begin{equation}
\mathcal{L}=\frac{\lambda _{0}}{2\lambda _{2}}\mu ^{-\varepsilon }(\partial
_{\mu }\chi )^{2}+\mu ^{-\varepsilon }\sum_{n=0}^{\infty }\!\!\!\!\!\!\left. %
\phantom{{a\over a}}\right. ^{\prime }\varepsilon ^{n}\frac{\lambda
_{2}^{n-2}}{\lambda _{0}^{2n-3}}[\partial ^{2n}]\overline{W}_{n}(\chi
,\varepsilon ),  \label{classi}
\end{equation}
where $\chi =\varphi \mu ^{\varepsilon /2}\sqrt{\lambda _{2}/\lambda _{0}}$
and I have suppressed the subscript B in $\overline{W}_{n}$. Defining 
\begin{equation}
\lambda _{2}=\frac{1}{M_{P}^{2}},\qquad M_{P\text{eff}}^{2}=M_{P}^{2}\lambda
_{0},\qquad \chi =\frac{\varphi \mu ^{\varepsilon /2}}{M_{P\text{eff}}},
\label{efficaci}
\end{equation}
the classical lagrangian (\ref{classi}) becomes 
\begin{equation}
\mathcal{L}=\frac{1}{2}\mu ^{-\varepsilon }M_{P\text{eff}}^{2}(\partial
_{\mu }\chi )^{2}+\mu ^{-\varepsilon }\sum_{n=0}^{\infty }\!\!\!\!\!\!\left. %
\phantom{{a\over a}}\right. ^{\prime }\frac{M_{P\text{eff}}^{4-2n}}{\lambda
_{0}^{n-1}}\varepsilon ^{n}[\partial ^{2n}]\overline{W}_{n}(\chi
,\varepsilon ).  \label{eva}
\end{equation}
I prove that (\ref{eva}) is equivalent to (\ref{vnform}), the negative
powers of $\lambda _{0}$ being effectively cancelled by the powers of $%
\varepsilon $.

The lagrangian (\ref{eva}) contains evanescent vertices, which can be traded
for higher-order local vertices. Consider an evanescent bare local operator $%
E_{\mathrm{B}}$. Being unsubtracted, its contribution is not negligible,
because the evanescence can simplify divergences and give finite
contributions. According to the theory of evanescent operators \cite{collins}%
, the renormalized operator $[E_{\mathrm{B}}]$ is equal to the sum of
non-evanescent renormalized local operators $[\mathrm{O}_{\mathrm{R}}]$ plus
truly evanescent local operators $[E_{\mathrm{R}}]$, 
\[
\lbrack E_{\mathrm{B}}]=[E_{\mathrm{R}}]+[\mathrm{O}_{\mathrm{R}}]. 
\]
``Truly'' evanescent operators are those that give vanishing contributions
in the physical limit ($\varepsilon \rightarrow 0$) of the 1PI action $%
\Gamma $ and can be consistently neglected. The finite operator $[\mathrm{O}%
_{\mathrm{R}}]$ is obtained simplifying evanescences with divergences. This
means that $[\mathrm{O}_{\mathrm{R}}]$ is higher-order in the expansion. Its
order can be computed as follows.

Consider a diagram $G$ with $L$ loops and $v_{n}$ vertices of type $%
\overline{W}_{n}$, constructed with (\ref{eva}). Write $V=\sum_{n}v_{v}$ and 
$\mathcal{E}=\sum_{n}nv_{n}$. Assume that $G$ has overall divergences (the
subdivergences being treated inductively in the usual way), namely $L+%
\mathcal{E}-V+1\geq 0$ (see (\ref{obaoba})). Look for the finite
contributions that are generated when the divergences of $G$ are simplified
by the powers of $\varepsilon $ attached to the vertices. Counting the
povers of $\lambda _{0}$ attached to the vertices, the desired contributions
are local and have the form 
\begin{equation}
\hbar ^{L+\mathcal{E}-V+1}\lambda _{0}(\hbar \lambda _{0})^{V-1-\mathcal{E}%
}M_{P\text{eff}}^{2-2L-2\mathcal{E}+2V}[\partial ^{2(L+\mathcal{E}%
-V+1)}]H(\chi ),  \label{finiteterms}
\end{equation}
The theorem of the appendix ensures that the maximal $\varepsilon $-pole of $%
G$ is $\leq V-1$. The evanescences attached to the vertices are $\geq 
\mathcal{E}$, because of (\ref{eva}). They can simplify and give finite
contributions only if $V-1-\mathcal{E}\geq 0$. Therefore the finite terms (%
\ref{finiteterms}) are of type (\ref{vnform}). Similarly, it is easy to
check that the divergent contributions have the form specified by (\ref
{conta})-(\ref{frefa}) with $\alpha \rightarrow \lambda _{0}$.

The evanescent vertices, together with appropriate finite and divergent
terms, can be reorganized in objects of type $[E_{\mathrm{R}}]$ and
neglected. The remaining finite vertices, together with appropriate
divergent terms, can be organized in objects of type $[\mathrm{O}_{\mathrm{R}%
}]$ and kept. The remaining divergent contributions are the net counterterms
that need to be subtracted away. Thus (\ref{eva}) generates a renormalized
lagrangian of the form 
\begin{equation}
\mathcal{L}_{\mathrm{R}}=\frac{\mu ^{-\varepsilon }}{2}M_{P\text{eff}%
}^{2}(\partial _{\mu }\chi )^{2}+\sum_{n=0}^{\infty }\mu ^{-\varepsilon
}\hbar ^{n}\lambda _{0}M_{P\text{eff}}^{4-2n}[\partial ^{2n}]\widetilde{W}_{%
\mathrm{R}n}(\hbar \lambda _{0},\chi ,\varepsilon ),  \label{undo}
\end{equation}
that can be converted into 
\begin{equation}
\mathcal{L}_{\mathrm{R}}=\frac{\mu ^{-\varepsilon }}{2}M_{P\text{eff}%
}^{2}(\partial _{\mu }\widetilde{\chi })^{2}+\sum_{n=0}^{\infty
}\!\!\!\!\!\!\left. \phantom{{a\over a}}\right. ^{\prime }\mu ^{-\varepsilon
}\hbar ^{n}\lambda _{0}M_{P\text{eff}}^{4-2n}[\partial ^{2n}]W_{\mathrm{R}%
n}(\hbar \lambda _{0},\widetilde{\chi },\varepsilon ).  \label{agreement2}
\end{equation}
with a field redefinition of the form 
\begin{equation}
\widetilde{\chi }=\chi +\hbar \lambda _{0}\sum_{n=0}^{\infty }\hbar ^{n}M_{P%
\text{eff}}^{-2n}[\partial ^{2n}]Q_{n}(\hbar \lambda _{0},\chi ,\varepsilon
).  \label{frefeps}
\end{equation}
The functions $\widetilde{W}_{\mathrm{R}n}$ and $W_{\mathrm{R}n}$ in (\ref
{undo}) and (\ref{agreement2}) have the form (\ref{conta}) with $\alpha
\rightarrow \lambda _{0}$, while the function $Q_{n}$ of (\ref{frefeps}) has
an expression similar to the $F_{n}$ of (\ref{frefa}), with the only
difference that in $Q_{n}$ the sum starts from $m=0$, to include finite
field redefinitions.

The argument just given provides another proof of (\ref{vnform}). The theory
is singular in the limit $\lambda _{0}\rightarrow 0$ at fixed $M_{P}$ and
trivial in the limit $\lambda _{0}\rightarrow 0$ at fixed $M_{P\text{eff}}$,
where the entire interacting sector disappears. These properties are
analogous to those found in ref.s \cite{pap2,pap3}, to which the reader is
referred for comparison.

\subsection{Calculation of the corrected classical potential}

I call ``corrected'' classical potential $W(\chi )$ the complete tree-level
potential. In the case of the $\varphi ^{4}+\varphi ^{6}$ theory, where 
\begin{equation}
W(\chi )=\frac{\chi ^{4}}{4!}+\frac{\chi ^{6}}{6!}+\mathcal{O}(\chi ^{8}),
\label{case}
\end{equation}
the $\mathcal{O}(\chi ^{8})$-corrections are determined self-consistently in
the way that I now describe, using one-loop results.

\begin{figure}[tbp]
\centerline{\epsfig{figure=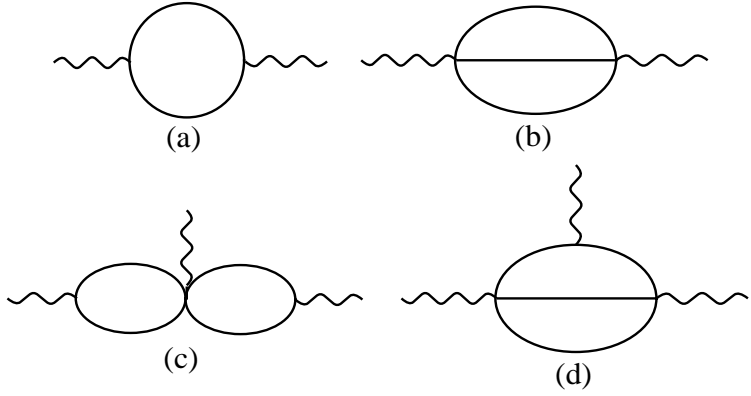,height=6cm,width=12cm}}
\caption{Relevant Feynman diagrams}
\end{figure}

\bigskip

\textbf{Differential equation for the corrected classical potential.} Start
from the lagrangian 
\begin{equation}
\mathcal{L}=\frac{1}{2}(\partial _{\mu }\varphi )^{2}+V(\varphi ,\lambda
_{0}\mu ^{\varepsilon },\lambda _{2}\mu ^{2\varepsilon },\varepsilon ),
\label{tho}
\end{equation}
where $V$ is analytic in $\varepsilon $, but can contain evanescent terms.
Using the background-field method, expand $\varphi $ as $\varphi _{0}+h$
inside (\ref{tho}), where $\varphi _{0}$ is the background field and $h$ is
the quantum fluctuation. For one-loop calculations it is sufficient to
expand up to $\mathcal{O}(h^{2})$. The order $\mathcal{O}(h)$ can be
dropped, since it does not contribute to the 1PI generating functional $%
\Gamma [\varphi _{0}]$. \ Thus the one-loop divergences are encoded in the
lagrangian 
\begin{equation}
\mathcal{L}_{h}=\frac{1}{2}(\partial _{\mu }h)^{2}+\frac{1}{2}h^{2}K,\qquad 
\text{where }K=V^{\prime \prime }(\varphi _{0},\lambda _{0}\mu ^{\varepsilon
},\lambda _{2}\mu ^{2\varepsilon },\varepsilon )\text{,}  \label{bach}
\end{equation}
and are given by diagram (a) of Fig.1. Here the prime denotes
differentiation with respect to $\varphi $. The counterterm is easily
evaluated: 
\[
\Delta \mathcal{L}_{h}=\frac{\hbar \mu ^{-\varepsilon }}{32\varepsilon \pi
^{2}}K^{2}, 
\]
so the one-loop renormalized lagrangian reads 
\begin{equation}
\mathcal{L}_{\mathrm{R}\text{ one-loop}}=\frac{1}{2}(\partial _{\mu }\varphi
)^{2}+V(\varphi ,\lambda _{0}\mu ^{\varepsilon },\lambda _{2}\mu
^{2\varepsilon },\varepsilon )+\frac{\hbar \mu ^{-\varepsilon }}{%
32\varepsilon \pi ^{2}}\left[ V^{\prime \prime }(\varphi ,\lambda _{0}\mu
^{\varepsilon },\lambda _{2}\mu ^{2\varepsilon },\varepsilon )\right] ^{2}.
\label{eller}
\end{equation}
Consistency with renormalization demands 
\begin{equation}
\mathcal{L}_{\mathrm{R}}=\mathcal{L}_{\mathrm{B}}=\frac{1}{2}(\partial _{\mu
}\varphi _{\mathrm{B}})^{2}+V_{\mathrm{B}}(\varphi _{\mathrm{B}},\lambda _{%
\mathrm{B}},\varepsilon ).  \label{custom}
\end{equation}
The bare and renormalized potentials have expansions: 
\begin{equation}
V_{\mathrm{B}}(\varphi _{\mathrm{B}},\lambda _{\mathrm{B}},\varepsilon )=%
\frac{\lambda _{0\mathrm{B}}^{3}}{\lambda _{2\mathrm{B}}^{2}}W_{\mathrm{B}%
}(\chi _{\mathrm{B}},\varepsilon ),\qquad V(\varphi ,\lambda ,\varepsilon
)=\mu ^{-\varepsilon }\frac{\lambda _{0}^{3}}{\lambda _{2}^{2}}%
\sum_{k=0}^{\infty }(\hbar \lambda _{0})^{k}W_{0,k}(\chi ,\varepsilon ).
\label{mimi}
\end{equation}
Both $W_{\mathrm{B}}(\chi _{\mathrm{B}},\varepsilon )$ and $W_{0,k}(\chi
,\varepsilon )$ are analytic in $\varepsilon $. Comparing the bare and
renormalized lagrangians it follows that $\varphi _{\mathrm{B}}=\varphi +%
\mathcal{O}(\hbar ^{2})$. The corrected classical potential is the function $%
W_{0,0}(\chi ,0)\equiv W(\chi )$.

The problem (\ref{custom}) is solved matching the bare and renormalized
lagrangians. Write $\lambda _{0\mathrm{B}}=\lambda _{0}\mu ^{\varepsilon
}Z_{0}$, $\lambda _{2\mathrm{B}}=\lambda _{2}\mu ^{2\varepsilon }Z_{2}$,
with $Z_{0}=1+a\hbar \lambda _{0}$, $Z_{2}=1+b\hbar \lambda _{0}$. Then the
matching gives immediately $W_{0,0}(\chi ,\varepsilon )=W_{\mathrm{B}}(\chi
,\varepsilon )$ and 
\begin{equation}
W_{0,1}(\chi ,\varepsilon )=-\frac{1}{32\varepsilon \pi ^{2}}\left[ \left(
W_{0,0}^{\prime \prime }(\chi ,\varepsilon )\right) ^{2}-32\pi
^{2}(3a-2b)W_{0,0}(\chi ,\varepsilon )+16\pi ^{2}(a-b)\chi W_{0,0}^{\prime
}(\chi ,\varepsilon )\right] ,  \label{sopra}
\end{equation}
where now the prime denotes differentiation with respect to $\chi $. Since $%
W_{0,1}(\chi ,\varepsilon )$ is analytic in $\varepsilon $ the pole in (\ref
{sopra}) cancels out. This gives the equation 
\begin{equation}
(3a-2b)W+\frac{\chi }{2}(-a+b)W^{\prime }=\frac{(W^{\prime \prime })^{2}}{%
32\varepsilon \pi ^{2}}\text{.}  \label{proto}
\end{equation}
Matching the powers $\chi ^{4}$ and $\chi ^{6}$, the one-loop
renormalization constants 
\[
Z_{0}=1+\frac{3\hbar \lambda _{0}}{16\varepsilon \pi ^{2}},\qquad Z_{2}=1+%
\frac{15\hbar \lambda _{0}}{16\varepsilon \pi ^{2}}, 
\]
are obtained, so the final equation for $W$ reads 
\begin{equation}
(W^{\prime \prime })^{2}+42W-12\chi W^{\prime }=0\text{.}  \label{newt}
\end{equation}
The power-counting solution is $W(\chi )=\chi ^{4}/4!$. Beyond
power-counting, the solution is uniquely specified by the initial conditions
(\ref{case}).

\bigskip

\textbf{Study of the corrected classical potential.} The non-linear
differential equation (\ref{newt}) is not of known types. Some exact
properties of the solution can be easily derived. Since $W$ is positive in
the neighborhood of $\chi =0$, let $[0,p)$ denote the maximal interval
containing $\chi =0$ where $W>0$. The equation (\ref{newt}) implies that $%
W^{\prime }>0$ in such an interval, so $p=\infty $: $W$ is everywhere
positive and monotonically increasing. We have the two inequalities 
\begin{equation}
12\chi W^{\prime }>(W^{\prime \prime })^{2},\qquad \frac{W^{\prime }}{W}\geq 
\frac{7}{2\chi }.\qquad  \label{uy}
\end{equation}
Integrating (\ref{uy}) between $\delta $ and $\chi >\delta $ we obtain the
estimates 
\[
\frac{W(\delta )}{\delta ^{7/2}}~\chi ^{7/2}\leq W(\chi )<\frac{\chi ^{4}}{3}%
+\frac{8c}{5\sqrt{3}}\chi ^{5/2}+c^{2}\chi +b, 
\]
where $b$ and $c$ are constants.

Finally, inserting a power-like behavior $W(\chi )\sim \beta \chi ^{\alpha }$
into the equation (\ref{newt}) for $\chi \rightarrow \infty $ it is
immediate to prove that the possible power-like behaviors at infinity are 
\[
W(\chi )\sim \beta \chi ^{7/2},\qquad W(\chi )\sim \frac{\chi ^{4}}{4!}, 
\]
with $\beta $ arbitrary.

Summarizing, the behavior of $W$ is regular, with a unique minimum in the
origin, and very strongly restricted. In particular, $W$ does not increase
faster than $\chi ^{4}$ at infinity, even if it is originated by a $\varphi
^{6}$ classical potential! Going through the arguments just derived, it is
easy to see that the same conclusions hold when $\lambda _{0}>0$, $\lambda
_{2}<0$: an unstable classical potential is turned into a stable corrected
classical potential.

The equation (\ref{newt}) can be studied around $\chi \sim 0$ with a series
expansion. The recursion relations for the coefficients of the expansion are
quadratic, see (\ref{genes}). I have calculated the first 400 coefficients
using Mathematica and checked empirically that the radius of convergence is
about 1. The first few terms of the expansion are 
\[
W(\chi )=\frac{\chi ^{4}}{4!}+\frac{\chi ^{6}}{6!}-\frac{\chi ^{8}}{1152}+%
\frac{7~\chi ^{10}}{20736}-\frac{203~\chi ^{12}}{1244160}+\frac{211~\chi
^{14}}{2488320}-\frac{28553~\chi ^{16}}{597196800}+\mathcal{O}(\chi ^{18}). 
\]
For $\chi \gtrsim 1$ the expansion badly diverges. With a Range-Kutta method
it is easy to numerically extend the solution beyond $\chi \sim 1$. I have
arrived at $\chi \sim 2.8$.

Thus, although the solution of (\ref{newt}) is regular everywhere, the
series expansion in powers of the fields diverges for large $\chi $. This
proves that the non-perturbative approach of this paper is useful in
concrete cases.

It should be remarked that the corrected classical potential $W(\chi )$ is
not a purely ``classical'' potential, because it knows about the quantum
theory: it can be determined only using one-loop knowledge. On the other
hand, $W(\chi )$ is not the quantum potential either, since the quantum
potential is contained in the 1PI generating functional $\Gamma $.
Summarizing, the quantization prescription for non-renormalizable theories
contains one step more than usual: the starting point is a classical
lagrangian, such as (\ref{claf}), which is generically unstable with respect
to renormalization; the first step is to determine a corrected classical
lagrangian that is stable under renormalization, such as (\ref{agreement2});
the last step is to compute the quantum action $\Gamma $.

\subsection{Two-loop calculations}

The two-loop relevant Feynman diagrams are ($b$), ($c$) and ($d$) of Fig. 1
and produce the counterterms 
\begin{equation}
\Delta \mathcal{L}_{\mathrm{R}\text{ two-loop}}=\frac{\hbar ^{2}\mu
^{-2\varepsilon }}{2\varepsilon (4\pi )^{4}}\left[ -\frac{1}{12}(\partial
_{\mu }V^{\prime \prime \prime })^{2}+\left( \frac{1}{\varepsilon }-\frac{1}{%
2}\right) V^{\prime \prime }(V^{\prime \prime \prime })^{2}+\frac{1}{%
\varepsilon }(V^{\prime \prime })^{2}V^{\prime \prime \prime \prime }\right]
,  \label{twoloops}
\end{equation}
that have to be added to the one-loop renormalized lagrangian $\mathcal{L}_{%
\mathrm{R}\text{ one-loop}}$ of (\ref{eller}). The first term of (\ref
{twoloops}) can be converted into a contribution to the potential by means
of a field redefinition. For this purpose, define the bare field 
\[
\varphi _{\mathrm{B}}=\varphi -\frac{\hbar ^{2}\mu ^{-2\varepsilon }}{%
24\varepsilon (4\pi )^{4}}\int_{0}^{\varphi }\left[ V^{\prime \prime \prime
\prime }(\varphi ^{\prime })\right] ^{2}\mathrm{d}\varphi ^{\prime }+%
\mathcal{O}(\hbar ^{3}). 
\]
Consistency with renormalization is expressed by the equation 
\begin{equation}
\mathcal{L}_{\mathrm{R}}=\mathcal{L}_{\mathrm{R}\text{ one-loop}}+\Delta 
\mathcal{L}_{\mathrm{R}\text{ two-loop}}=\mathcal{L}_{\mathrm{B}}=\frac{1}{2}%
(\partial _{\mu }\varphi _{\mathrm{B}})^{2}+V_{\mathrm{B}}(\varphi _{\mathrm{%
B}},\lambda _{\mathrm{B}},\varepsilon ),  \label{probo}
\end{equation}
where the bare and renormalized potentials have the expansions (\ref{mimi}).
The function $W_{0,1}(\chi ,0)\equiv Y(\chi )$ is the one-loop contribution
to the corrected classical potential and its initial condition is $Y(\chi )=%
\mathcal{O}(\chi ^{8})$. Indeed, the powers of $\varphi ^{4}$ and $\varphi
^{6}$ in $V(\varphi ,\lambda ,0)$ are multiplied by the independent
couplings $\lambda _{0}$ and $\lambda _{2}$, so no $\hbar $-corrections are
necessary in front of $\varphi ^{4}$ and $\varphi ^{6}$. This means that $%
W_{0,n}(\chi ,0)=\mathcal{O}(\chi ^{8})$ for $n>1$.

The problem (\ref{probo}) is solved matching the bare and renormalized
lagrangians. This matching gives immediately the functions $W_{0,i}(\chi
,\varepsilon )$, $i=0,1,2$, in terms of $W_{\mathrm{B}}(\chi _{\mathrm{B}%
},\varepsilon )$. Then, using the fact that $W_{0,i}$ are analytic in $%
\varepsilon $, differential or integro-differential equations for $W(\chi )$
and $Y(\chi )$ are obtained. Matching the powers $\chi ^{4}$ and $\chi ^{6}$
in such equations the two-loop renormalization constants 
\begin{equation}
Z_{0}=1+\frac{3\lambda _{0}}{16\varepsilon \pi ^{2}}+\left( \frac{9}{%
\varepsilon ^{2}}-\frac{17}{6\varepsilon }\right) \frac{\lambda _{0}^{2}}{%
(4\pi )^{4}},\qquad Z_{2}=1+\frac{15\lambda _{0}}{16\varepsilon \pi ^{2}}%
+\left( \frac{135}{\varepsilon ^{2}}-\frac{427}{12\varepsilon }\right) \frac{%
\lambda _{0}^{2}}{(4\pi )^{4}},  \label{rui}
\end{equation}
are obtained. At two loops, the procedure just outlined gives two equations,
one for the double pole and one for the simple pole. The equation due to the
double pole is equivalent to (\ref{newt}). Indeed, the one-loop simple pole
and the two-loop double pole are related to each other by the
renormalization group \cite{collins}. The equation given by the simple
two-loop pole is

\[
18Y-6\chi Y^{\prime }+W^{\prime \prime }Y^{\prime \prime }=\frac{1}{192\pi
^{2}}\left( 1504W-393\chi W^{\prime }+6W^{\prime \prime }(W^{\prime \prime
\prime })^{3}-W^{\prime }\int_{0}^{\chi }\left[ W^{\prime \prime \prime
\prime }(\chi ^{\prime })\right] ^{2}\mathrm{d}\chi ^{\prime }\right) 
\]
and has a unique solution with the initial condition $Y(\chi )=\mathcal{O}%
(\chi ^{8})$, once $W(\chi )$ is known. The first few coefficients of the
solution are

\[
Y(\chi )=\frac{\chi ^{8}}{46080\pi ^{2}}\left[ 43-\frac{401}{30}\chi ^{2}+%
\frac{883}{72}\chi ^{4}-\frac{201497}{23760}\chi ^{6}+\mathcal{O}(\chi
^{8}).\right] 
\]

\subsection{Inclusion of the mass term}

With a mass term, 
\[
\mathcal{L}_{\text{cl}}=\frac{1}{2}(\partial _{\mu }\varphi )^{2}+V_{\text{cl%
}}(\varphi ,\lambda ),\qquad V_{\text{cl}}(\varphi ,\lambda )=\frac{m^{2}}{2}%
\varphi ^{2}+\frac{\lambda _{0}}{4!}\varphi ^{4}+\frac{\lambda _{2}}{6!}%
\varphi ^{6}, 
\]
the dimensionless combinations are 
\[
\frac{\varphi _{\mathrm{B}}^{2}\lambda _{2\mathrm{B}}}{\lambda _{0\mathrm{B}}%
},\qquad m_{\mathrm{B}}^{2}\frac{\lambda _{2\mathrm{B}}}{\lambda _{0\mathrm{B%
}}^{2}},\qquad \lambda _{0\mathrm{B}}^{2(\varepsilon -1)}\lambda _{2\mathrm{B%
}}^{-\varepsilon }. 
\]
The corrected classical potential can depend only on the first two. Defining 
$\tau =m^{2}\lambda _{2}/\lambda _{0}^{2}$, we have 
\[
\qquad V(\varphi ,\lambda )=\frac{\lambda _{0}^{3}}{\lambda _{2}^{2}}W(\chi
,\tau ),\qquad W(\chi ,\tau )=\tau \frac{\chi ^{2}}{2!}+\frac{\chi ^{4}}{4!}+%
\frac{\chi ^{6}}{6!}+\mathcal{O}(\chi ^{8})\text{.} 
\]
Proceeding with the methods explained in the previous subsections, it is
possible to derive a partial differential equation for $W(\chi ,\tau )$,
with coefficients related to the one-loop renormalization constants. The
equations can be solved perturbatively in $\tau $, starting from the
solution $W(\chi )$ found before. The result is 
\begin{eqnarray*}
W(\chi ,\tau ) &=&W(\chi )+\tau \left( \frac{\chi ^{2}}{2!}-\frac{\chi ^{8}}{%
288}+\frac{7~\chi ^{10}}{4320}-\frac{791~\chi ^{12}}{518400}+\mathcal{O}%
(\chi ^{14})\right) + \\
&&\tau ^{2}\left( -\frac{17~\chi ^{8}}{684}+\frac{106~\chi ^{10}}{12825}+%
\mathcal{O}(\chi ^{12})\right) +\tau ^{3}\left( -\frac{41437~\chi ^{8}}{%
198360}+\mathcal{O}(\chi ^{10})\right) +\mathcal{O}(\tau ^{4}\chi ^{8}).
\end{eqnarray*}
The renormalization constants of the couplings are 
\begin{eqnarray*}
m_{\mathrm{B}}^{2} &=&m^{2}\left( 1+\frac{\lambda _{0}}{16\varepsilon \pi
^{2}}\right) ,\qquad \qquad \lambda _{0\mathrm{B}}=\lambda _{0}\left( 1+%
\frac{(\tau +3)\lambda _{0}}{16\varepsilon \pi ^{2}}\right) , \\
\lambda _{2\mathrm{B}} &=&\lambda _{2}\left[ 1+\left( 15-35\tau -140\tau
^{2}-\frac{19040~\tau ^{3}}{19}+\mathcal{O}(\tau ^{4})\right) \frac{\lambda
_{0}}{16\varepsilon \pi ^{2}}\right] .
\end{eqnarray*}

\section{The $\varphi ^{4}+\varphi ^{m+4}$ theory in $D=4$: conditions for a
meaningful reduction of couplings}

\setcounter{equation}{0}

In this section I consider a more general class of irrelevant deformations,
those with 
\begin{equation}
V_{\text{cl}}(\varphi ,\lambda )=\frac{\lambda _{0}}{4!}\varphi ^{4}+\frac{%
\lambda _{m}}{(m+4)!}\varphi ^{m+4},  \label{irre}
\end{equation}
where $m$ is not necessarily integer, but positive. The dimensionality of $%
\lambda _{m\mathrm{B}}$ is 
\[
\lbrack \lambda _{m\,\mathrm{B}}]=-m+\frac{\varepsilon }{2}\left( m+2\right)
, 
\]
so the only acceptable dimensionless combination is 
\[
\frac{\varphi _{\mathrm{B}}^{m}\lambda _{m\,\mathrm{B}}}{\lambda _{0\mathrm{B%
}}}\equiv \chi _{\mathrm{B}}^{m}, 
\]
and the corrected classical potential has the form 
\begin{equation}
V(\varphi ,\lambda )=\left( \frac{\lambda _{0}^{m+4}}{\lambda _{m}^{4}}%
\right) ^{1/m}W(\chi ).  \label{evident}
\end{equation}
Defining $\lambda _{m\mathrm{B}}=\lambda _{m}\mu ^{(m+2)\varepsilon /2}Z_{m}$%
, $Z_{m}=1+b_{m}\hbar \lambda _{0}$, and using the strategy described in the
previous section, the differential equation 
\[
\frac{(m+4)a-4b_{m}}{m}W+\frac{b_{m}-a}{m}\chi W^{\prime }=\frac{(W^{\prime
\prime })^{2}}{32\varepsilon \pi ^{2}} 
\]
is obtained. The values of $a$ and $b_{m}$ are calculated matching the
orders $\chi ^{4}$ and $\chi ^{m}$ of the solution. This gives the
renormalization constant of $\lambda _{0}$ and the one of $\lambda _{m}$: 
\begin{equation}
Z_{m}=1+\frac{(m+4)(m+3)}{32\varepsilon \pi ^{2}}\hbar \lambda _{0}.
\label{abbare}
\end{equation}
The final equation for $W$ reads 
\[
m(W^{\prime \prime })^{2}+2(m+4)(2m+3)W-(m+6)(m+1)\chi W^{\prime }=0\text{%
,\qquad }W(\chi )=\frac{\chi ^{4}}{4!}+\frac{\chi ^{m+4}}{(m+4)!}+\mathcal{O}%
(\chi ^{8}). 
\]
The study of the solution proceeds as before. It is easily found that $W$ is
positive, monotonically increasing and its large-$\chi $ behavior is bounded
by 
\[
\beta \chi ^{\frac{2(m+4)(2m+3)}{(m+6)(m+1)}}\lesssim W(\chi )\lesssim \frac{%
\chi ^{4}}{4!}. 
\]

Expand $W(\chi )$ in powers of $\chi $, 
\begin{equation}
W(\chi )=\sum_{n=0}^{\infty }c_{nm}\chi ^{nm+4},\qquad c_{0}=\frac{1}{4!}%
,\qquad c_{m}=\frac{1}{(m+4)!}.  \label{evi}
\end{equation}
Observe that the expansion of $W$ is consistent with the fact that only
non-negative integer powers of $\lambda _{m}$ and integer powers of $\lambda
_{0}$ appear in $V(\varphi ,\lambda )$, see (\ref{evident}). The
coefficients $c_{nm}$, $n>1$, are determined by the recursion relations 
\begin{equation}
c_{nm}=-\frac{1}{(n-1)(nm^{2}-6)}%
\sum_{k=1}^{n-1}c_{km}c_{(n-k)m}(km+3)(km+4)\left( (n-k)m+3\right) \left(
(n-k)m+4\right) .  \label{genes}
\end{equation}
The first few terms of the expansion are 
\begin{eqnarray*}
W(\chi ) &=&\frac{\chi ^{4}}{4!}+\frac{\chi ^{m+4}}{(m+4)!}-\frac{1}{2}\frac{%
\chi ^{2m+4}}{(m^{2}-3)[(m+2)!]^{2}}+\frac{1}{3}\frac{(m+2)(2m+3)~\chi
^{3m+4}}{(m^{2}-2)(m^{2}-3)[(m+2)!]^{3}}+ \\
&&-\frac{1}{6}\frac{(m+2)(2m+3)(8m^{4}+21m^{3}-8m^{2}-56m-36)~\chi ^{4m+4}}{%
(m^{2}-2)(2m^{2}-3)(m^{2}-3)^{2}[(m+2)!]^{4}}+\mathcal{O}(\chi ^{5m+4}).
\end{eqnarray*}

\bigskip

\textbf{Reduction of couplings.} I\ show that formula (\ref{genes}) solves
the one-loop RG-consistency equations (\ref{rg}). Write 
\begin{equation}
V(\varphi ,\lambda )=\sum_{n=0}^{\infty }\frac{\lambda _{nm}}{(nm+4)!}%
\varphi ^{nm+4}.  \label{caso}
\end{equation}
The reduction relations to this order can be read from (\ref{evident}) and (%
\ref{evi}): 
\[
\qquad \frac{\lambda _{nm}}{(nm+4)!}=a_{nm}(\hbar \lambda _{0})\frac{\lambda
_{m}^{n}}{\lambda _{0}^{n-1}}, 
\]
with $a_{nm}(0)=c_{nm}$. Reading the one-loop anomalous dimensions from (\ref
{abbare}), 
\[
\gamma _{nm}^{(1)}=\frac{(nm+4)(nm+3)}{32\pi ^{2}}, 
\]
and using $\beta _{0}^{(1)}=3/(16\pi ^{2})$, the existence conditions (\ref
{conditions}) become 
\begin{equation}
r_{n,m}=\frac{1}{6}(n-1)(nm^{2}-6)\notin \mathbb{N}\text{.}  \label{gy}
\end{equation}

Consider first the case of integer $m$, or, more generally, rational $m^{2}$%
. It is clear from (\ref{gy}) that infinitely many $n$s violate the
existence condition. Each violation implies the introduction of a new
parameter at order $r_{n,m}\in \mathbb{N}$ in $\lambda _{0}$. In particular,
the corrected classical potential is uniquely determined if $r_{n,m}$ is not
zero, in agreement with (\ref{genes}).

For $m=2$, which is the theory $\varphi ^{4}+\varphi ^{6}$, no value $n>1$
gives $r_{n,2}=0,1$, so no new parameter appears at the tree and one-loop
levels, in agreement with the results of the previous section, which gave
unique functions $W(\chi )$ and $Y(\chi )$. The first integer values of $%
r_{n,2}$ are 2,5,15,22,40$\ldots $ The first new parameter appears at two
loops.

In section 4 it was proved that the introduction of new parameters and the
renormalization mixing do not modify (\ref{gy}). At the lowest order the
renormalization mixing (calculated in the undeformed theory $\mathcal{R}$)
can be non-triangular only among operators that have the same number of
derivatives. Consider the operators O$_{p,q}$ of (\ref{opa}) with fixed $p$
and arbitrary $q$. Simple combinatorics shows that the one-loop anomalous
dimensions, therefore also $r_{n,\ell }$ and $r_{n,k,\ell }^{IJ}$, grow
quadratically with the number $q$ of legs. Therefore, the four-derivative
operators (\ref{affo}), which have a trivial one-loop renormalization
mixing, provide invertibility conditions similar to (\ref{gy}). Their
violations are expected to be infinitely many, but the new parameters are
introduced at orders that grow with $q$. Operators with more than four
derivatives, e.g. (\ref{affa}), have a non-trivial one-loop renormalization
mixing, so their invertibility conditions are violated much more rarely,
again at orders growing with $q$.

In conclusion, the complete theory is expected to work precisely as
illustrated by the one-loop results. The number of new couplings grows
together with the order of the expansion. Every finite order is described by
a finite number of independent couplings. A relatively small number of
couplings is sufficient to make physical predictions up to very high orders.
A reliable estimate of the number of new couplings and the orders at which
they appear can be obtained counting the violations of (\ref{gy}), so
calculations up to the fortieth order in the theory $\varphi ^{4}+\varphi
^{6}$ with operators of arbitrarily high dimensionalities are expected to
require about ten independent couplings. Non-renormalizable theories
formulated in the naive way contain infinitely many independent couplings
already at the tree level, thus the simplification and the gain in
predictive power are enormous.

The conditions (\ref{gy}) are always fulfilled when $m^{2}$ is irrational.
Also the invertibility conditions associated with the four-derivative
operators (\ref{affo}) are quadratic in $m$, so they are expected to be
violated at most by a finite set of $q$s. Thus, non-analytic theories with
irrational $m^{2}$ are expected to be renormalizable with a finite number of
independent couplings.

Summarizing, in general potentials with irrational $m^{2}$ can be quantized
consistently with a finite number of independent couplings, while rational
potentials contain a number of independent couplings that grows sporadically
with the order of the perturbative expansion. In the next section I consider
the theory $\varphi ^{4}+\varphi ^{5}$ in four dimensions, where a new
independent coupling appears in the corrected classical potential.

\section{The $\varphi ^{4}+\varphi ^{5}$ theory in $D=4$: appearance of new
parameters}

\setcounter{equation}{0}

The theory $\varphi ^{4}+\varphi ^{5}$ in four dimensions has $m=1$ and $%
r_{6,1}=0$. The equation for $W(\chi )$, which reads 
\begin{equation}
(W^{\prime \prime })^{2}+50W-14\chi W^{\prime }=0,  \label{ey}
\end{equation}
does not admit a power series solution, because the formula (\ref{genes})
has a singular denominator for $n=6$, $m=1$. A new independent coupling $%
\overline{\lambda }_{6}$ has to be introduced in front of $\varphi ^{10}$.
It is convenient to define a dimensionless running parameter $\xi $. Its
beta function is calculated from (\ref{eller}) with the usual procedure: 
\[
\overline{\lambda }_{6}=\xi \frac{\lambda _{1}^{6}}{\lambda _{0}^{5}},\qquad
\beta _{\xi }=\frac{5915}{972}\frac{\hbar \lambda _{0}}{32\pi ^{2}}. 
\]
The modified equation for $W(\chi ,\xi )$ reads 
\begin{equation}
(W^{\prime \prime })^{2}+50W-14\chi W^{\prime }=\frac{5915}{972}\frac{%
\partial W}{\partial \xi },  \label{ryo}
\end{equation}
whose solution has an expansion 
\begin{eqnarray}
W(\chi ,\xi ) &=&\frac{\chi ^{4}}{24}+\frac{\chi ^{5}}{120}+\frac{\chi ^{6}}{%
144}+\frac{5}{432}\chi ^{7}+\frac{355}{10368}\chi ^{8}+\frac{6545}{31104}%
\chi ^{9}+\xi \,\chi ^{10}-\frac{450275}{69984}\chi ^{11}  \label{finestra}
\\
&&-5\,\xi \,\chi ^{11}+\frac{67399525}{3359232}\chi ^{12}+\frac{125}{12}\xi
\,\chi ^{12}+\mathcal{O}(\chi ^{13}).  \nonumber
\end{eqnarray}
Once the new parameter $\xi $ is introduced, the solution $W(\chi ,\xi )$ of
(\ref{ryo}) is uniquely determined.

For completeness, I report an alternative way to solve the problem, which
does not modify equation (\ref{ey}). The solution of (\ref{ey}) contains a
new finite parameter $\zeta $ ($\beta _{\zeta }=0$), but is not analytic in
the fields: 
\[
W(\chi ,\zeta )=\left. W(\chi )\right| _{9}-\frac{1183}{972}\chi ^{10}\ln
(\zeta \,\chi )-\frac{502327}{69984}\chi ^{11}+\frac{5915}{972}\chi ^{11}\ln
(\zeta \,\chi )+\mathcal{O}(\chi ^{12}), 
\]
where $\left. W(\chi )\right| _{9}$ are the terms $\chi ^{4}$-$\chi ^{9}$ of
(\ref{finestra}).

\section{Conclusions}

\setcounter{equation}{0}

Are power-counting renormalizable theories more ``fundamental'' than the
other theories? In the light of the results obtained so far, the reasons
usually advocated to privilege the power-counting renormalizable theories
over the non-renormalizable ones appear to be weak. Predictivity and
calculability are not peculiar features of renormalizable theories. Here and
in ref.s \cite{pap2,pap3} I have shown that calculability belongs to a much
larger class of models, that include also power-counting non-renormalizable
theories, which can be formulated in a perturbative framework, often with a
finite or reduced set of independent couplings, without assuming knowledge
about the ultraviolet limit. The so-formulated theories have a non-trivial
predictive content. Therefore, if power-counting renormalizable theories are
viewed as fundamental, then there are several other theories that have to be
considered equally fundamental. On the other hand, our present notion of
fundamental theory might still be imperfect. The research pursued here might
contribute to find the right definition.

The techniques to reduce the number of independent couplings consistently
with renormalization are useful tools to classify the non-renormalizable
interactions, although they cannot select which ones are switched on and off
in nature. The key issue is the number of independent couplings that are
necessary to subtract the divergences. Some insight in this problem is
provided by the construction of finite and quasi-finite non-renormalizable
theories \cite{pap2,pap3}, which are irrelevant deformations of interacting
conformal field theories. Their divergences are renormalized away just with
field redefinitions (finiteness), plus, eventually, a finite number of
independent renormalization constants (quasi-finiteness). The
non-renormalizable theories studied in this paper are irrelevant
deformations of running renormalizable theories. They admit a simplified
renormalization structure, where, for example, terms with $2n$ derivatives
of the fields, $n>1$, do not appear before the $n^{\text{th}}$ loop. The
dependence on the fields, although not on their derivatives, can be treated
non-perturbatively. A suitable perturbative expansion can be defined, such
that each order of the quantum action is calculable with a finite number of
steps. I have studied the predictive power of these theories, in particular
under which conditions the divergences are renormalized with a reduced,
eventually finite, number of independent couplings. In the simplest
(analytic) non-renormalizable theories the number of parameters grows
sporadically with the order of the expansion. It becomes infinite in the
complete theory, but the growth is so slow that a reasonably small number of
parameters is sufficient to make predictions up to very high orders. Most
non-analytic theories can be renormalized with a finite number of couplings
in a strict sense.

\vskip 25truept \noindent {\Large \textbf{Acknowledgments}}

\vskip 15truept \noindent

I am grateful to M. Matone and A. Kitaev for useful correspondence on the
properties of the differential equation (\ref{newt}), E. Vicari and M.
Campostrini for discussions about the numerical analysis of the solution,
and M. Mintchev for discussions on irrational theories.

\vskip 25truept \noindent {\Large \textbf{A\ \ Appendix: a theorem on the
maximal logarithmic divergences of diagrams}}

\vskip 15truept

\renewcommand{\theequation}{A.\arabic{equation}} \setcounter{equation}{0}

\noindent Here I prove a general result that is used in the paper. Recall
that the UV divergences are calculated treating the mass term, if present,
as a (two-leg) vertex. The propagator is just $1/k^{2}$, so tadpoles vanish.

\bigskip

\textbf{Theorem}. The maximal pole of a diagram with $V$ vertices and $L$
loops is at most of order $m(V-1,L)\equiv \min (V-1,L)$.

\textbf{Proof}. I prove the statement inductively in $V$ and, for fixed $V$,
inductively in $L$. The diagrams with $V=1$ and arbitrary $L$ are tadpoles,
which vanish identically and therefore satisfy the theorem. Suppose that the
statement is true for $V<\overline{V}$, $\overline{V}>1$, and arbitrary $L$.
Consider diagrams with $\overline{V}$ vertices. Clearly for $L=1$ the
maximal divergence is $1/\varepsilon $, so the theorem is satisfied. Proceed
inductively in $L$, i.e. suppose that the theorem is satisfied by the
diagrams with $\overline{V}$ vertices and $L<\overline{L}$ loops, and
consider the diagrams $G_{\overline{V},\overline{L}}$ that have $\overline{V}
$ vertices and $\overline{L}$ loops. If $G_{\overline{V},\overline{L}}$ has
no subdivergence, its divergence is at most a simple pole. Higher-order
poles are related to the subdivergences of $G_{\overline{V},\overline{L}}$
and can be classified replacing the subdiagrams with their counterterms.
Consider the subdiagrams $\gamma _{v,l}$ of $G_{\overline{V},\overline{L}}$,
with $l$ loops and $v$ vertices. Clearly, $1\leq l<\overline{L}$ and $1\leq
v\leq \overline{V}$. By the inductive hypothesis, the maximal divergence of $%
\gamma _{v,l}$ is a pole of order $m(v-1,l)$. Contract the subdiagram $%
\gamma _{v,l}$ to a point and multiply by $1/\varepsilon ^{m(v-1,l)}$. A
diagram with $\overline{V}-v+1\leq $ $\overline{V}$ vertices and $\overline{L%
}-l<\overline{L}$ loops is obtained, whose maximal divergence, taking into
account of the factor $1/\varepsilon ^{m(v-1,l)}$, is at most a pole of
order $m(v-1,l)+m(\overline{V}-v,\overline{L}-l)$. The inequality 
\[
m(v-1,l)+m(\overline{V}-v,\overline{L}-l)\leq m(\overline{V}-1,\overline{L}%
), 
\]
which can be derived case-by-case, proves that the maximal divergence of $G_{%
\overline{V},\overline{L}}$ associated with $\gamma _{v,l}$ satisfies the
theorem. Since this is true for every subdiagram $\gamma _{v,l}$, the
theorem follows for $G_{\overline{V},\overline{L}}$. By induction, the
theorem follows for every diagram.

\end{document}